\documentclass{aa}
\usepackage{txfonts}
\usepackage{natbib}
\usepackage{graphicx}
\usepackage{psfig}

\def\nustar{{\em NuSTAR}}

\def \srcon{IGR J17503-2636}
\def \srcoff{IGR J17507-2647}

\def \igrtre{IGR J17503-2636}
\def \igrsette{IGR J17507-2647}

\def\xmm{{\em XMM--Newton}}
\def\sw{{\em Swift}}
\def\swift{{\em Swift}}
\def\inte{{\em INTEGRAL}}

\def\chandra{{\em Chandra}}
\def\nustar{{\em NuSTAR}}

\def\gaia{{\em Gaia}}

\def\approxgt{\mathrel{\hbox{\rlap{\lower.55ex \hbox {$\sim$}}
        \kern-.3em \raise.4ex \hbox{$>$}}}}
\def\approxlt{\mathrel{\hbox{\rlap{\lower.55ex \hbox {$\sim$}}
        \kern-.3em \raise.4ex \hbox{$<$}}}}

\def\pdot {$\dot P_{\rm spin}$}

\def\flux {\mbox{erg cm$^{-2}$ s$^{-1}$}}

\def\nh{$N_{\rm H}$}
\def\ltsima{$\; \buildrel < \over \sim \;$}
\def\lsim{\lower.5ex\hbox{\ltsima}}
\def\gtsima{$\; \buildrel > \over \sim \;$}
\def\gsim{\lower.5ex\hbox{\gtsima}}

\def\hcm {\hbox {\ifmmode $ atom cm$^{-2}\else atom cm$^{-2}$\fi}}

\def\arcsec {\hbox{$^{\prime\prime}$}}

\def \apj {ApJ}
\def \aj {AJ}
\def \apjl {ApJL}
\def \apjs {ApJS}
\def \aap {A\&A}
\def \aaps {A\&AS}

\def \mnras {MNRAS}

\def \nar {New Astronomy Reviews}

\def \ssr {Space Science Reviews}

\newcommand{\be}{\begin{equation}}

\newcommand{\ee}{\end{equation}}
\newcommand{\ergs}{{~\rm erg\, s^{-1}}}

\newcommand{\msun}{{~M_{\odot}}}

\begin{document}

  \title{New interpretation of the two hard X-ray sources \\
  IGR J17503-2636 and IGR J17507-2647
}

\author{L.~Sidoli\inst{1}, V.~Sguera\inst{2},  P.~Esposito\inst{3}, R.~Sathyaprakash\inst{3}, G.~Ponti\inst{4,5}, S.~Mondal\inst{4}, A.J.~Bird\inst{6}  
}
\institute{$^1$ INAF, Istituto di Astrofisica Spaziale e Fisica Cosmica, via A.\ Corti 12, I-20133 Milano, Italy \\
$^2$ INAF, Osservatorio di Astrofisica e Scienza dello Spazio, Via P. Gobetti 101, I-40129 Bologna, Italy \\
$^3$ Scuola Universitaria Superiore IUSS Pavia, Piazza della Vittoria 15, I-27100, Pavia, Italy  \\
$^4$ INAF, Osservatorio Astronomico di Brera, via E. Bianchi 46, I-23807 Merate (LC), Italy \\
$^5$ Max-Planck-Institut f{\"u}r extraterrestrische Physik, Giessenbachstrasse, D-85748, Garching, Germany \\
$^6$ School of Physics and Astronomy, University of Southampton, Southampton SO17 1BJ, UK 
}

\offprints{L. Sidoli, lara.sidoli@inaf.it}

\date{Received 18 July 2024 / Accepted 21 November 2024}

\authorrunning{L. Sidoli et al.}

\titlerunning{Two hard X-ray sources revised}

\abstract{
  We report on the results of X-ray observations (\xmm, INTEGRAL and \swift) 
  of two hard X-ray sources, IGR\,J17503-2636 and  IGR\,J17507-2647, whose nature is not fully elucidated in the literature.
  Three \xmm\ observations covered the field of IGR\,J17503-2636, in 2020 and twice in 2023.
  The analysis of the two \xmm\ observations performed in September 2023, six days apart,  did not detect \igrtre, allowing us 
  to pose the most stringent 3$\sigma$ upper limit
  on the source flux to date ($\sim$9.5$\times$10$^{-14}$~\flux, 2-10 keV, flux corrected for absorption).
  This value implies that the amplitude of the X-ray flux variability exceeds a factor of $\sim$2100, compared with the discovery outburst in 2018.
A candidate X-ray periodicity at 0.335397(3) seconds  has been barely detected (significance of $\sim3.8\sigma$) from \igrtre\ with \xmm\  
(pulsed fraction of (10 $\pm$ 1)\%). 
  The new data, put into the context of previous literature, 
  allow us to propose a new  classification of IGR\,J17503-2636 as a symbiotic X-ray binary, rather than 
  a candidate supergiant fast X-ray transient.
  IGR\,J17507-2647 was formerly reported below 10 keV only during \textit{Chandra} observations performed in 2009. 
  We report here on two \xmm\ observations that serendipitously covered the source field in 2020 and in 2023, 
  finding a  stable X-ray emission, both in X-ray flux and spectral shape. The long-term, persistent X-ray emission has also been probed by several \sw/XRT short observations  and by  INTEGRAL data spanning several years.
  We have detected an iron line in the emission (with centroid energy in the range of 6.3-6.6 keV), never reported before in the IGR\,J17507-2647 spectrum. The source properties favor the identification with a cataclysmic variable.
  \keywords{stars: neutron - X-rays: binaries - pulsars: individual:  IGR\,J17503-2636, IGR\,J17507-2647, IGR\,J17505-2644}
}
\maketitle

        \section{Introduction\label{intro}}

        The catalogs of hard X-ray sources discovered by the INTEGRAL satellite (IGRs) include many unclassified objects, or sources for which only a tentative identification has been proposed, deserving confirmation. 
        In particular, we have been searching the IBIS catalogs \citep{Bird2006, Bird2007, Bird2010, Bird2016} for IGRs that have been suggested to be X-ray binaries (XRBs) and, in particular, candidate supergiant fast X-ray transients (SFXTs), thanks to follow-up observations at softer energies (below 10 keV). To obtain a more in-depth view, we have searched  the public high-energy archives for unpublished observations covering the source sky regions. 
        Here, we focus on two IGRs, \igrtre\ and \igrsette, falling within three \xmm\ observations. 
        Their small angular distance ($\sim$10 arcmin) implies that they can be observed  within the same \xmm\ pointing. Therefore, we have decided to discuss them together in this paper.  
      
        The X-ray transient \igrtre\ was discovered on August 11, 2018, during observations with Jem-X on board the
        INTEGRAL satellite \citep{Chenevez2018}. It showed fluxes of 2$\times10^{-10}$\,\flux\ (3-10 keV)
        and 1.9$\times10^{-10}$\,\flux\ (10-25 keV). However, it was not detected by IBIS/ISGRI on board INTEGRAL
        above 20 keV, with an upper limit (3$\sigma$) of 10 mCrab (20-40 keV).
        A \sw/XRT observation performed two days later refined the sky position \citep{Chenevez2018}.
        The spectrum was highly absorbed (\nh=(13$\pm{5})$$\times$10$^{22}$ cm$^{-2}$) and showed
        an intrinsic flux of 
        1.4$^{+0.3} _{-0.2}$$\times$10$^{-10}$\,\flux\ (2-10 keV) when fit with a power law model (with a photon index of $\Gamma$=0.5$\pm{0.6}$; \cite{Ferrigno2019}).
        The source sky position was refined at the sub-arcsec level
        thanks to a \chandra\ observation performed on August 23, 2018 \citep{Chakrabarty2018a, Chakrabarty2018b}.
        During this observation, the source was faint (0.010(3) counts s$^{-1}$, HRC-S),
        with an absorbed flux of 2.9$\times10^{-12}$\,\flux\ (0.2-10 keV), assuming
        a power law model (photon index of $\Gamma$=2 and \nh=1$\times$10$^{23}$ cm$^{-2}$).
        $NuSTAR$, $NICER$, and $Swift$/XRT follow-up observations performed during
        the discovery outburst were reported by \citet{Ferrigno2019, Ferrigno2022}, showing a long-term
        X-ray flux variability by a factor of about 300.
        A flux variability on much shorter timescales of a few seconds to
        a few thousand seconds was observed (0.3-80 keV, \citealt{Ferrigno2019}).
        Some spectral variability is also present,
        as well as a changing absorbing column density, likely local to the source \citep{Ferrigno2019}.
        A fluorescent iron line was observed in $NuSTAR$ data, together
        with a hint of a cyclotron feature at 20 keV.
        No X-ray pulsations were found \citep{Ferrigno2019}.
        \citet{Masetti2018} identified the near-infrared (NIR)
        counterpart of \igrtre\ (Z$>$19.9 mag, Y=17.90 mag, J=14.23 mag, H=11.79 mag, K$_{s}$=10.65 mag) and
        concluded that the NIR photometry is consistent with an heavily reddened OB (super)giant located
       at a distance of about 10 kpc.
       This led them to suggest that \igrtre\ is a high-mass X-ray binary (HMXB), belonging to the subclass of 
       the SFXTs (\citealt{Sguera2005, Sguera2006, Negueruela2006}).
       \citet{McCollum2018} modeled the spectral energy distribution (SED)
       of this object, adding more data and extending measurements to the mid-infrared (MIR) region
       ({\em Spitzer} and WISE data).
       The best fit of the SED fitting implies a cool giant (M4-6 III star) with an IR excess
       (at $\lambda$ $>$ 12 $\mu$m).
       This result questions the conclusion drawn by \citet{Masetti2018} from the NIR photometry, but it has never been discussed further in the literature.

        \igrsette\ (also known as  IGR\,J17505-2644) 
        is a hard X-ray source whose behavior is largely unknown. 
         It was discovered by INTEGRAL  and reported for the first time
         in the IBIS catalogs by \citealt{Bird2007} (with the name \igrsette) 
         and by \citet{Krivonos2007} as IGR\,J17505-2644. 
         In this paper, we adopt the source name \igrsette. 
         It is also listed in the two subsequently published  IBIS catalogs  (\citealt{Bird2010}; \citealt{Bird2016}), and is consistently best detected 
         in the energy band 18-60 keV as a persistent source. The typical average flux is 1.1$\pm{0.1}$ mCrab in both energy bands 20--40 keV and 40--100 keV \citep{Bird2016}.
        No further investigation  was performed until a \chandra\ (ACIS-I)
        observation in February 2009, which pinpointed the soft X-ray counterpart
        CXOU\,J175039.4-264436 \citep{Tomsick2009}.
        The \chandra\ spectrum was fit with a power law model with a photon index, $\Gamma$=0.44$^{+0.84} _{-0.72}$,
        and a high absorbing column density, \nh=13.4$(^{+7.8} _{-5.5})\times10^{22}$ cm$^{-2}$.
        The flux, corrected for absorption, was 4.5($^{+1.9}_{-0.7}$)$\times10^{-12}$\,\flux\ (0.3-10 keV).
        The source does not have any 2MASS, DENIS, USNO-B1.0, or USNO-A2.0 counterpart, leading \citet{Tomsick2009}
        to suggest that the high absorption is probably interstellar, indicative of a large distance,
        near the Galactic center. In this case, the  X-ray luminosity is
        $\sim$4$\times10^{34}$\,erg\,s$^{-1}$, favoring an XRB
        rather than a cataclysmic variable (CV) \citep{Tomsick2009}.
        A second \chandra\ observation (HRC-I) targeted on the source was performed a few months later, in August 2009 \citep{Zolotukhin2011}, resulting in sky coordinates consistent with the one reported by \citealt{Tomsick2009}.
        \citet{Zolotukhin2011} marginally detected a NIR counterpart in the
        UKIDSS-DR3 Galactic Plane Survey \citep{Lucas2008} 
        inside the \chandra/HRC-I error circle, with K=18.5$\pm{0.4}$~mag,
        while in J and H filter only upper limits were derived (J$>$20.3 mag, 
        H$>19.3$~mag). 
        Assuming that this NIR counterpart is mainly due to emission from an accretion disk in a low-mass X-ray binary (LMXB), these authors estimate an orbital period of $\sim$390 hours. However, we note that Zolotukhin et al. assume that the source is a LMXB simply based on the sky position in the direction of the Galactic bulge. 
        No \gaia\ counterparts are available for either IGR source, so the sources' distances are unknown. In this paper, we report on unpublished, archival X--ray data covering the sky position of both sources, allowing us to obtain a more sensitive spectral investigation of their X-ray emission and to reconsider the source nature proposed previously.

\section{Observation and data reduction}
\label{data}

The data analyzed here consist of  \xmm, \sw\ and INTEGRAL observations
covering the sky position of \igrtre\ and \igrsette. Given the small angular separation ($\sim$10 arcmin)
between them, on two occasions  a single \xmm\ pointing  covered both sources, as well as some \sw/XRT observations.
In the following sections, we discuss 
\xmm\  (Sect.\,\ref{sect:xmmdata}), \sw\  (Sect.\,\ref{sect:swiftdata}), 
and INTEGRAL observations (Sect.\,\ref{sect:intedata}).

\subsection{\xmm}
\label{sect:xmmdata}

Three unpublished \xmm\ \citep{Jansen2001} observations are relevant here:
an observation performed in 2020 (ObsID 0844101101) targeted at \igrtre,
and two shorter observations that were part of a survey of the
Galactic plane (ObsID 0932190801 and 0932191001). In the following subsections, we discuss them separately, given the different instrument modes and issues depending on the off-axis position.

The log of the observations is reported 
in Table\,\ref{tab:xmmlog}, while in Table\,\ref{tab:srcxmm} 
we briefly summarize when the two sources have been detected or not, 
for clarity.
Given the faintness of the sources, only EPIC \citep{Struder2001, Turner2001} is relevant.

\begin{figure}
 \includegraphics[width=8.7cm, angle=0]{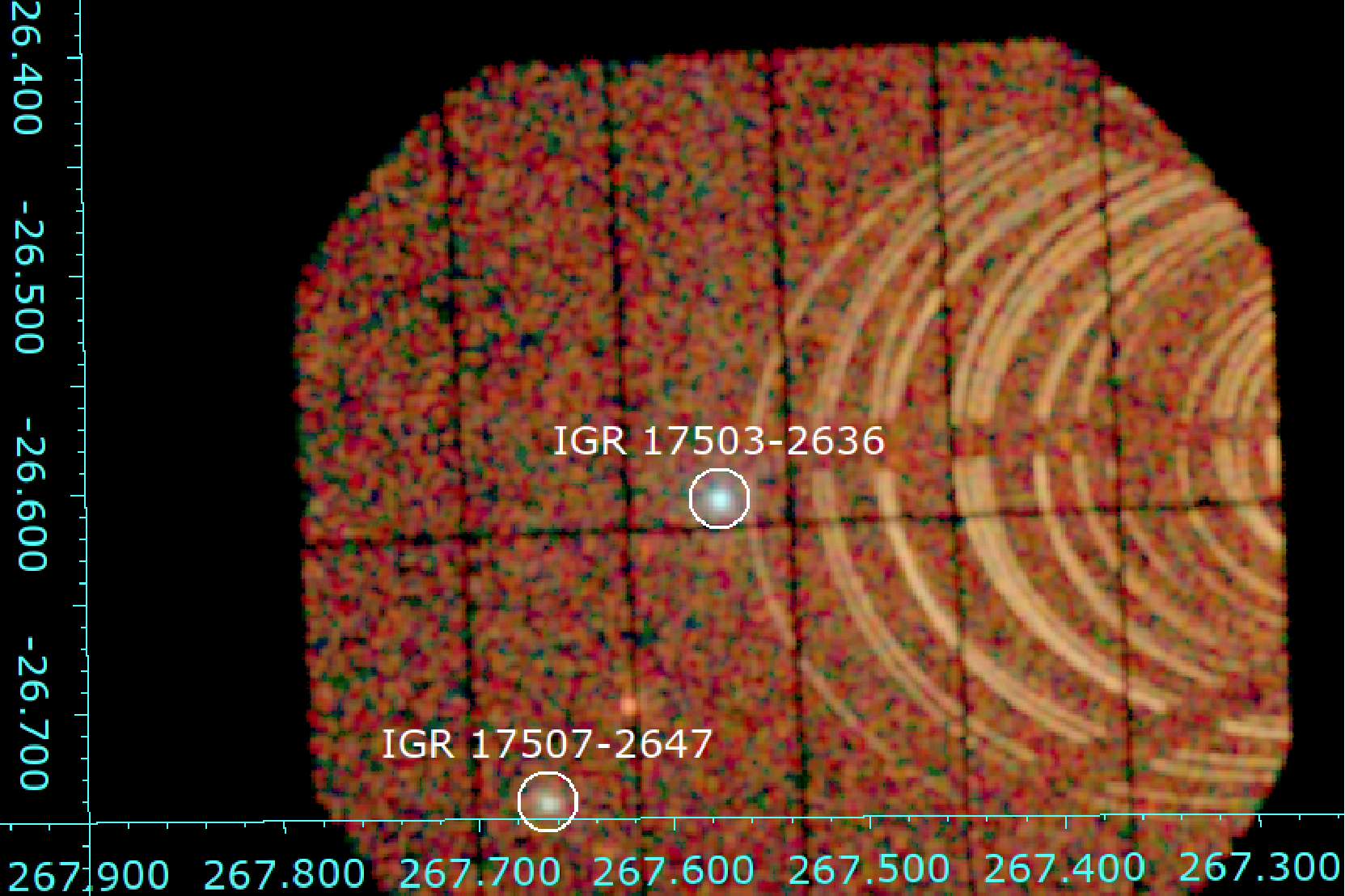} 

 \includegraphics[width=8.7cm, angle=0]{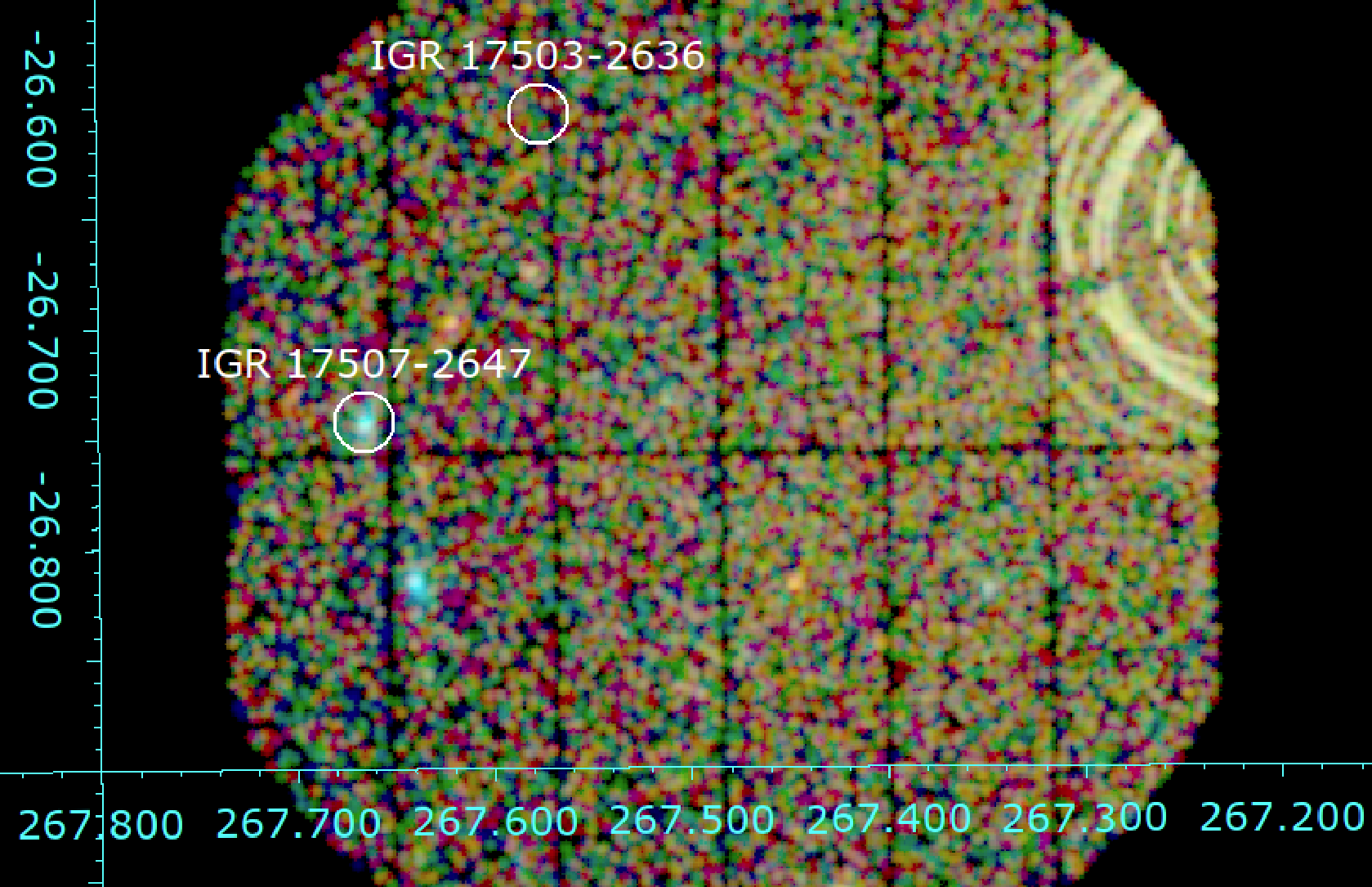} 

  \includegraphics[width=8.7cm, angle=0]{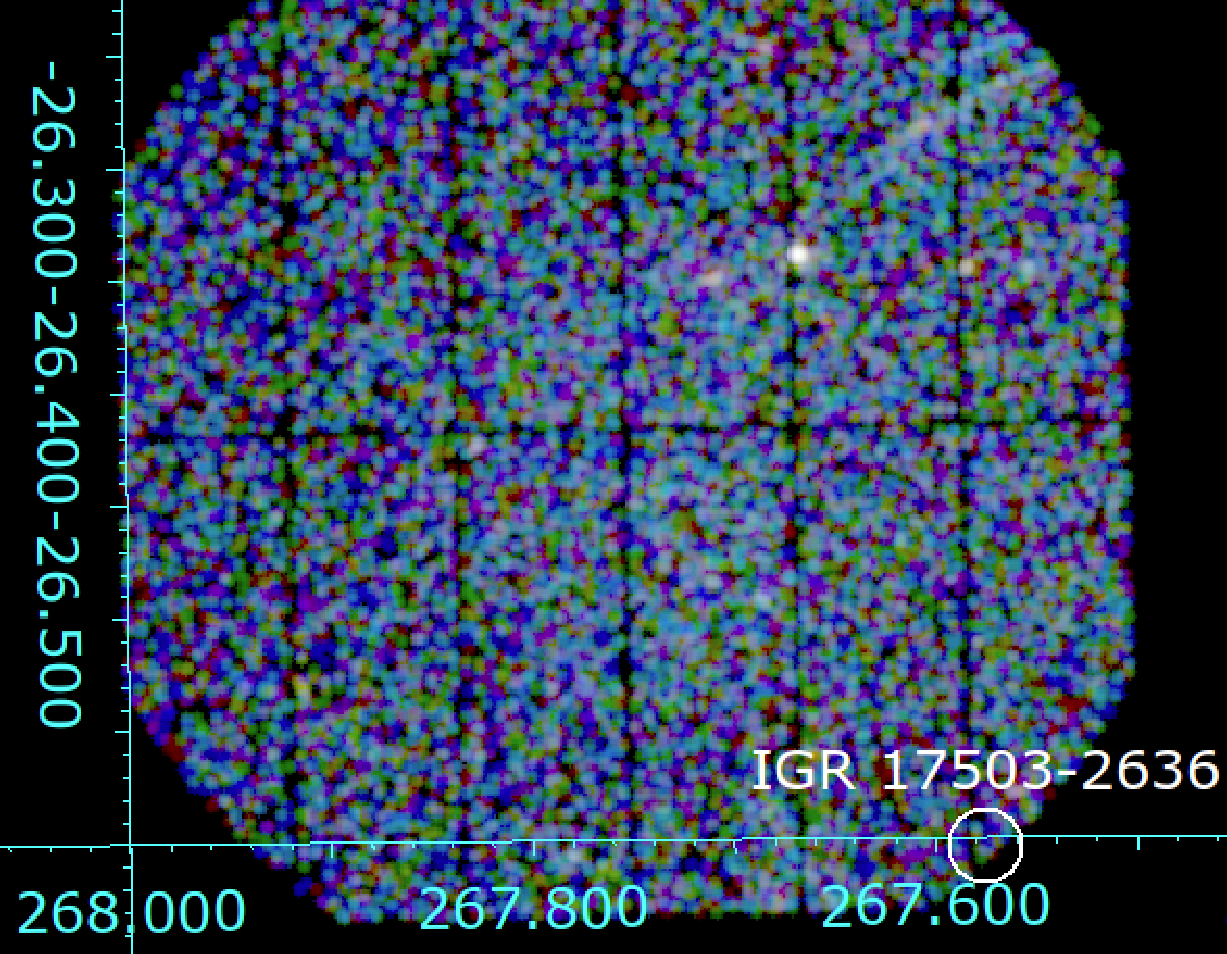}
 \caption{EPIC pn images of the three \xmm\ observations. From top to bottom: ObsID 0844101101, 0932190801, and 0932191001 (0.3-12 keV). 
 The black circles (1 arcmin radius) mark the position of \igrtre\ and \igrsette. Arches on the right of each image are due to stray light contamination by a bright source located outside the FoV, probably GX\,3+1. Equatorial coordinates (J2000) are reported.  
}
\label{fig:ima}
\end{figure}

In Fig.~\ref{fig:ima}, we show the three EPIC pn images, together
with the sky position of the sources of interest. 
Some stray light contamination produced by a bright source outside the field of view (FoV), probably the bright source GX\,3+1, is present.
When this  contamination lies close to the sky position of the two sources, we mention this issue explicitly.

\begin{table*}
        \centering
        \caption{\xmm\ observations (MOS2 exposure data are reported).}
        \label{tab:xmmlog}
        \begin{tabular}{lccc}
                \hline
                ObsID    &   Start time  &   End time       &  Nominal Exp.     \\
                         &       (TT)   &   (TT)              &   (ks)    \\
\hline
0844101101  &   2020-04-03 at 19:15:26      &  2020-04-04 at 02:18:03   &  25.3    \\
0932190801  &   2023-09-04 at 11:57:34      &  2023-09-04 at 15:11:00   &  11.6     \\
0932191001  &   2023-09-10 at 03:07:44      &  2023-09-10 at 05:56:10   &  10.0    \\     
\hline
        \end{tabular}
\end{table*}

\begin{table}
        \centering
        \caption{Sources in the \xmm\ observations.}
        \label{tab:srcxmm}
        \begin{tabular}{lcc}
                \hline
ObsID    &   \igrtre\  &   \igrsette\      \\
\hline
0844101101  &  detected            &   detected  \\
0932190801  &  undetected          &   detected     \\
0932191001  &  undetected          & sky position not covered      \\            
\hline
        \end{tabular}
\end{table}

\xmm\ data were reprocessed using the version 21
of the Science Analysis Software (SAS), with standard procedures.
The response and ancillary matrices
were generated with {\sc rmfgen} and {\sc arfgen} available in the SAS.
The background was extracted from source-free regions close
to the source position, and was not contaminated by external stray light.

EPIC spectra were simultaneously fit using {\sc xspec} \citep{Arnaud1996}.
We included multiplicative constants to take into account calibration uncertainties. 
 We fixed at 1 the multiplicative constant for EPIC pn spectra, while we left  the multiplicative constant of EPIC MOS1 and MOS2 free to vary during the fitting procedure. After fitting, we checked that  these constant factors were always compatible with 1, for all spectra analyzed here.
When fitting the spectra, the absorption model {\sc TBabs} was used, with
photoelectric absorption cross sections of \cite{Verner1996} and the interstellar abundances of \cite{Wilms2000}. 
The spectra were rebinned to have at least 20 counts per bin, to apply the $\chi^{2}$ statistics. 
All  spectral uncertainties are given at a 90\% confidence level, for
a single interesting parameter \citep{Avni1976}, using the tool {\sc error}\footnote{https://heasarc.gsfc.nasa.gov/xanadu/xspec/manual/XspecManual.html}  in {\sc xspec}  \citep{Arnaud1996}.

The uncertainties on the unabsorbed X-ray fluxes (UFs) were derived using the {\sc cflux} tool in {\sc xspec}.
When the sources were undetected, upper limits on the net count rates (2-12 keV) were estimated  using the {\sc eupper} tool, available in the SAS. 
Exposure maps were produced using the SAS tool {\sc eexpmap}.

\subsubsection{ObsID 0844101101}

The archival \xmm\  \citep{Jansen2001} observation ID 0844101101,
performed in April 2020, was targeted at \srcon. 
The EPIC FoV also covered the position of \srcoff.
EPIC pn operated in full frame mode, adopting the medium filter. MOS1 operated in a small window and MOS2 in timing mode.
All three cameras were available for data reduction and product extraction of \srcon,
whereas for the off-axis source \igrsette\ we could extract products
only for the pn and MOS2 (falling on an external charge-coupled device, CCD, operated in imaging mode). 
In fact, the CCD of the MOS1 camera covering the position of the off-axis source was not operational. 
  
Source light curves and spectra were extracted 
from circular regions centered on the source emission,
with a 15\arcsec\ radius for \srcoff\ and 30\arcsec\ for \srcon, 
and selecting a pattern from 0 to 4 (EPIC pn), and from 0 to 12 (MOS).
Background spectra were obtained from larger regions
(with a radius of 60\arcsec) offset from the source position, but on the same CCD, 
and away from the stray light contamination.
For the MOS2 data in timing mode, the source spectrum was extracted
selecting counts with RAWX coordinates in the range of 295-315, 
with a background spectrum extracted from the strip 255-265 (RAWX pixels).

\subsubsection{ObsID 0932190801}

This EPIC observation is part of the Survey of the Galactic plane
and it serendipitously covered the sky position of \igrtre\ and \igrsette. 
However, only \igrsette\ was detected and it was bright enough for spectroscopy (reported in Sect.~\ref{sect:xmmrestwo}), while for \igrtre\ we have calculated the upper limits on its X-ray flux (Sect.~\ref{sect:xmmrestwo}).

\subsubsection{ObsID  0932191001}

This EPIC observation is part of the Survey of the Galactic plane 
and it serendipitously covered the sky position of  IGR J17503-2636 only.
Since in this observation \igrtre\ was undetected, we have calculated  upper limits on its X-ray flux, which are reported  in Sect.\ref{sect:xmmresthree}.

\subsection{Neil Gehrels \sw\ Observatory}
\label{sect:swiftdata}

The log of the \textit{Neil Gehrels }\sw\  observations (\sw\ hereafter) analyzed here is reported in Table~\ref{tab:swlog}.
The XRT observations were reduced using {\sc xrtpipeline}, adopting standard procedures. 
Source detection and eventual estimation of the 3$\sigma$ upper limits to the
count rates (0.3-10 keV) was performed using {\sc ximage} software on images and exposure maps extracted in the energy range 0.3-10 keV. 
The tools {\sc detect} (to detect the sources) and {\sc sosta} for a proper evaluation of the source count rate (or upper limit) using a local background were used to obtain the source intensity corrected for the point spread function, sampling dead time, and vignetting.

\begin{table}
        \centering
        \caption{ \sw\ observations analyzed here}
        \label{tab:swlog}
        \begin{tabular}{lcccc}
                \hline
ObsID    &   Start time  &     Exp.    & IGR  \\
         &       (MJD)   &      (ks)  &  covered     \\
\hline
10807001 &  58343.8193    & 1.0    &   both  \\
10980001 &  58576.7122    & 5.2    &   both  \\
10980002 &  58583.4879    & 3.4    &   both  \\
10980003 &  58590.5288    & 4.9    &   both \\
10980004 &  58599.7178    & 0.1    &   both   \\
10980005 &  58611.7754    & 4.0    &   both   \\    
10980006 &  58618.0983    & 4.9    &   both   \\  
10980007 &  58643.7934    & 4.7    &   both   \\ 
43694001 &  56130.2847    & 0.6    &   both  \\  
43701001 &  56148.2576    & 0.6    &   \igrtre   \\   
48022002 &  56197.6333    & 0.7    &   \igrsette \\    
48022003 &  56201.4243    & 0.7    &   both  \\    
48022004 &  56209.9805    & 1.4    &   both  \\   
48022007 &  56224.2083    & 1.7    &   both  \\   
88805001 &  58353.1923    & 1.9    &   both  \\ 
\hline
        \end{tabular}
\end{table}

\subsection{\inte}
\label{sect:intedata}

The temporal behavior of both \igrsette\ and \igrtre\  has been  investigated  above 20 keV with  the ISGRI detector (Lebrun et al. 2003), which is the lower energy layer of the IBIS coded mask telescope (Ubertini et al. 2003) on board INTEGRAL (Winkler et al. 2003). 
INTEGRAL  observations are divided into short pointings (science windows, ScWs) whose  typical duration is $\sim$ 2,000 seconds. 
We considered only ScWs during which the sources were  within 12$^\circ$ 
of the center of the IBIS/ISGRI FoV.  A 12$^\circ$ limit is generally applied because the off-axis response of 
IBIS/ISGRI is not well modeled at large off-axis angles and in combination with the telescope dithering 
(or the movement of the source within the FoV) it may introduce a significant systematic error in the measurement 
of the source fluxes.  IBIS/ISGRI images for each ScW were generated in the energy band
18-60 keV using  the INTEGRAL Science Data Centre (ISDC) offline scientific  analysis software (OSA) version 11.2. 
ISGRI count rates at the position of the  sources 
were extracted from all individual images 
to produce their long-term light curve on the ScW timescale.  
Since \igrsette\ and \igrtre\  have a significantly different temporal behavior  above 20 keV (i.e., persistent and transient, respectively),  we adopted two different methods for our temporal investigation.  

\igrtre\ is a transient source discovered by INTEGRAL during 
revolution 1,986 (August 2018); hence, we used public IBIS/ISGRI observations from revolution 30 to 1,985 (i.e., from approximately January 2003 to August 2018) to search for any possible transient hard X-ray  activity before its discovery.
The dataset consists of 19,662 ScWs  where  the source was within 12$^\circ$ of the center of the IBIS/ISGRI FoV. The corresponding effective exposure time is equal to $\sim$28 Ms. In particular, we used the bursticity method developed by  Bird  et al. (2010, 2016) to search for any transient hard X-ray  activity  in a systematic way.  Such a method optimizes the  source detection timescale by scanning the IBIS/ISGRI light curve with a variable time window to search for the best source  significance value on timescales ranging from 0.5 days to 
weeks, months, or years. Then, the exact duration, time interval, and energy band over which the source significance is maximized are recorded (results in Sect.~\ref{sect:interestre}).

\igrsette\ is a persistent source best detected in the energy band 18-60 keV, as is reported in the latest  published INTEGRAL IBIS catalog, which considered satellite revolutions up to 1,000 \citep{Bird2016}. Hence, we used public IBIS/ISGRI observations from revolution 30 to 1,000 -- that is, from approximately January 2003  to December 2010 -- to produce  the source long-term light curve (18-60 keV) on the ScW timescale. The dataset  consists of 9,262 ScWs  where  the source was within 12$^\circ$ of the center of the instrument FoV. The effective exposure time is equal to $\sim$12 Ms.  The produced long-term light curve   was investigated to highlight any possible flaring or strongly variable behavior of the source on top of its persistent nature (results in Sect.~\ref{sect:interesette}).

The summary of the INTEGRAL observations analyzed here is reported in Table~\ref{tab:intelog} for both sources.
The interested reader can use this information to download the list of the about 30,000 ScWs from public archives, like HEASARC.\footnote{https://heasarc.gsfc.nasa.gov/}

\begin{table}
\caption{INTEGRAL observations analyzed  (range of the revolution number dataset, effective exposure, start time of the first and last ScW from  the  rev dataset during which the source was in the INTEGRAL FoV}
\label{tab:intelog}
\footnotesize
\begin{tabular}{c  c c c}
\hline
\hline
   Rev dataset    &     exp   &  start ScW    &  end ScW   \\
                            &     (Ms)         &    (UTC)                  &   (UTC)   \\
\hline
\hline
 &     &   &      \\
 \multicolumn{4}{c}{IGR J17507$-$2647}  \\
 &     &   &      \\
 30-1000 &  12   & 2003-02-28 03:44  & 2010-10-28 06:41  \\
\hline
 &     &   &      \\
 \multicolumn{4}{c}{IGR J17503$-$2633}  \\
 &     &   &      \\
 30-1985 &  28    & 2003-02-28 03:44  &  2018-04-23 09:38 \\
 \hline
  \hline
  \end{tabular}
\begin{list}{}{}
\item
\end{list}
\end{table}

\section{Results}
\label{sect:res}

In the following subsections, the results for both sources,
obtained with \xmm, \sw\ , and INTEGRAL satellites are reported.

\subsection{\xmm}
\label{sect:xmmres}

We report here the EPIC spectral and temporal results, separately for each \xmm\ observation, for clarity.

\subsubsection{\xmm\ ObsID 0844101101}
\label{sect:xmmresone}

\begin{figure*}
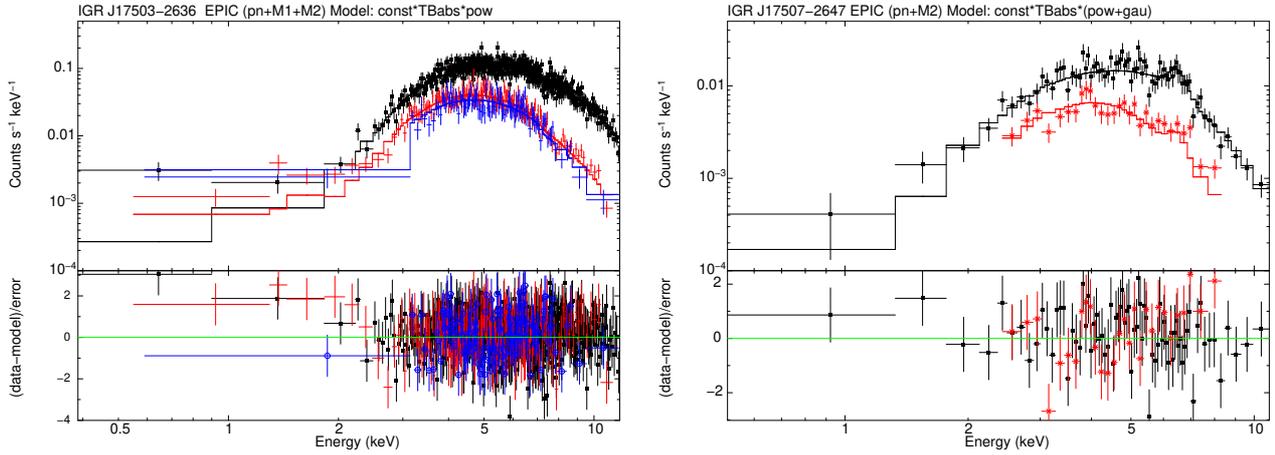

 \includegraphics[width=6.0cm, angle=-90]{ldadelchi_src_onaxis_pnm12_pow_blu.ps} 
  \includegraphics[width=6.0cm, angle=-90]{src_offaxis_pnm2_ldadelchi_pow_gau_final_new.ps}
\caption{Best fit of the EPIC counts spectra extracted from the whole observation (Obs ID 0844101101):  
\srcon\ is reported on the left (EPIC pn in black, MOS1 in red, and MOS2 in  blue) 
and  \srcoff\ is on the right (EPIC pn in black, MOS2 in red).   
Lower panels show the residuals in units of standard deviation. 
Spectral parameters are listed in Table~\ref{tab:spec}.
}
    \label{fig:spec}
\end{figure*}

The results of the time-averaged spectroscopy 
are reported in Table~\ref{tab:spec} for both sources. 
An absorbed power law is already a good fit to the spectrum of \srcon, while
the addition of a Gaussian line at $\sim$6.4 keV is needed to account for the emission of \igrsette.
Count spectra, together with the residuals, are shown in Fig.~\ref{fig:spec}. 
Both sources show a flat power law continuum (with a photon index, $\Gamma$, around 1.0) 
and a very high absorbing column density.
The significance of the 6.4 keV line in the spectrum of  \igrsette\ was tested running 10$^{6}$ simulations with the tool {\sc simftest} in {\sc xspec}, obtaining a probability of 3.70$\times10^{-4}$
that data are consistent with the model without the emission line component.

In order to investigate the eventual contribution to 
the iron emission line in \igrsette\ by highly ionized iron (6.7 and 6.9 keV lines), we added three Gaussian narrow lines (width fixed at 0) to the absorbed power law continuum. The Gaussian line energies were fixed at 6.4, 6.7, and 6.9 keV. 
The fit resulted in no evidence for the presence of ionized iron, as is reported in Table~\ref{tab:spec}, where  upper limits (90\% c.l.) on these line fluxes are reported. 
The lack of any ionized iron lines (6.7 and 6.9 keV) is confirmed by the fact that fitting the \igrsette\ spectrum with an {\sc apec} model 
(i.e., emission from a collisionally ionized gas, in {\sc xspec})
resulted in a worse fit ($\chi^{2}$=135.33 for 101 dof) 
than the simple power law, with a temperature of kT$>$40 keV.

\begin{table*}
        \centering
        \caption{Time-averaged EPIC spectroscopy (Obs ID 0844101101)  of \igrtre\ (second column) and \igrsette\ (last four columns) .
          The flux (UF) has been corrected for absorption (1-10 keV). 
        d$_{10\,kpc}$ is the source distance in units of 10 kpc. 
        }
        \label{tab:spec}
        \begin{tabular}{l|c|cccc}
                \hline
        Param             &  \srcon    &             \multicolumn{4}{c}{\srcoff}     \cr
                \hline
\nh\  (10$^{22}$~cm$^{-2}$)  &    23$\pm{1}$               &   11$\pm{2}$            &    12$\pm{2}$           & 12$\pm{2}$         &   15$\pm{1}$       \\                
Power law $\Gamma$           &   1.03$\pm{0.08}$           & 0.86$^{+0.24} _{-0.23}$ & 0.98$^{+0.27} _{-0.25}$ &  0.97$^{+0.26} _{-0.24}$   &  $-$  \\
UF (10$^{-11}$~\flux)        &   1.73$^{+0.09} _{-0.07}$   & 0.41$^{+0.04} _{-0.03}$ & 0.41$^{+0.05} _{-0.04}$ &  0.41$^{+0.05} _{-0.04}$  &  0.48$^{+0.03} _{-0.03}$  \\
kT$_{apec}$  (keV)           &        $-$                   &     $-$                &   $-$                    &     $-$          &     $>$40   \\
E$_{line}$ (keV)             &    $-$                      & $-$                     &   6.49$^{+0.12} _{-0.13}$ & 6.4 fixed    &  $-$  \\
line width  (keV)             &     $-$                    &         $-$            &   0.27$^{+0.15} _{-0.14}$  &  0.0 fixed   &  $-$    \\
norm. ($10^{-6}$ ph~cm$^{-2}$~s$^{-1}$)&     $-$           &         $-$            &    18$^{+8} _{-7}$         &  7.9$^{+3.6} _{-3.6}$    &  $-$     \\
EW (eV)                       &     $-$                    &         $-$            &   400$\pm{200}$            &   160$\pm{80}$        &  $-$     \\
E$_{line}$ (keV)             &    $-$                      & $-$                     &  $-$                      &  6.7 fixed          &     $-$       \\
line width  (keV)             &     $-$                    &         $-$            &  $-$                       &   0.0 fixed         & $-$   \\
norm. ($10^{-6}$ ph~cm$^{-2}$~s$^{-1}$)&     $-$           &         $-$            &   $-$                      &   $<$7.3      & $-$   \\
EW (eV)                       &     $-$                    &         $-$            &  $-$                       &   $<$130      &  $-$    \\
E$_{line}$ (keV)             &    $-$                      & $-$                     &  $-$                       &    6.9 fixed         & $-$  \\
line width  (keV)             &     $-$                    &         $-$            &   $-$                       &   0.0 fixed          &  $-$    \\
norm. ($10^{-6}$ ph~cm$^{-2}$~s$^{-1}$)&     $-$           &         $-$            &   $-$                      &   $<$7.7 & $-$   \\
EW (eV)                       &     $-$                    &         $-$            &  $-$                       &    $<$180    &    $-$   \\

L$_{\rm X}$ (erg~s$^{-1}$) & 2.1$\times$10$^{35} d^2_{10\,kpc}$  & 4.9$\times$10$^{34} d^2_{10\,kpc}$  & 4.9$\times$10$^{34} d^2_{10\,kpc}$   & 4.9$\times$10$^{34} d^2_{10\,kpc}$ &  5.7$\times$10$^{34} d^2_{10\,kpc}$  \\
$\chi^{2}$ (dof)             &     924.34 (886)            &      127.11 (101)      &   105.55 (98)            &    106.87 (98)     &   135.33 (101) \\
                \hline
        \end{tabular}
\end{table*}

\begin{figure*}
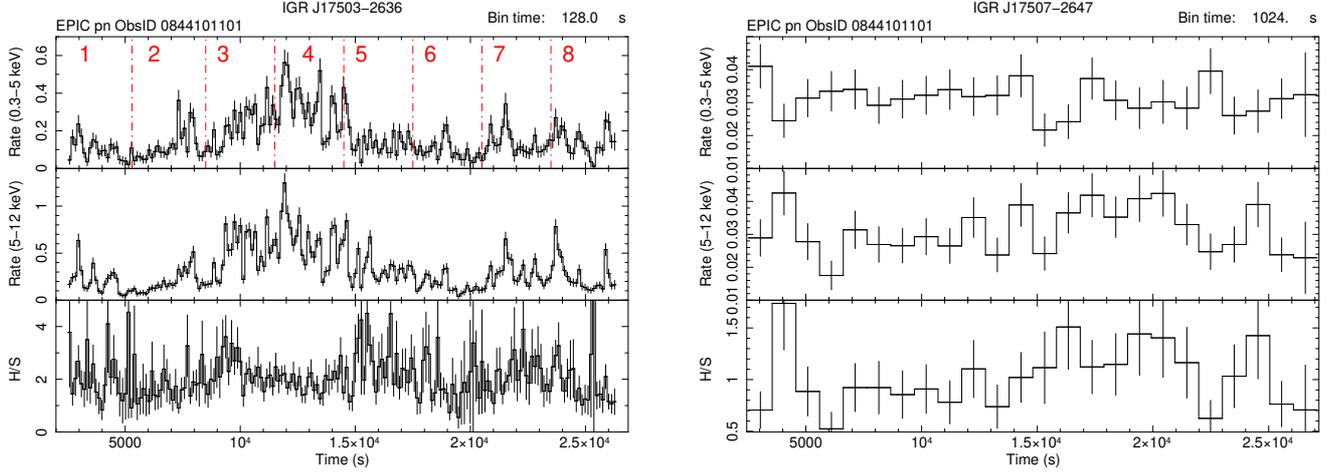

 \includegraphics[width=6.3cm, angle=-90]{net_lc_hr_igr17503_pn_04_0844101101_vertlines.ps} 
  \includegraphics[width=6.3cm, angle=-90]{net_igr17507_pn_p04_300_5000_12000_1024s.ps}
  \caption{EPIC pn, background-subtracted light curves (Obs ID 0844101101) of both sources (\igrtre\ on the left, \igrsette\ on the right),
    extracted in two energy ranges (above and below 5 keV),
    together with their hardness ratios in the lowest panels. 
 Vertical, dash-dotted red lines indicate the eight time intervals used to perform the time-selected spectroscopy (Table~\ref{tab:ontimesel}).
}
    \label{fig:lc}
\end{figure*}

\begin{figure}
 \includegraphics[width=6.5cm, angle=-90]{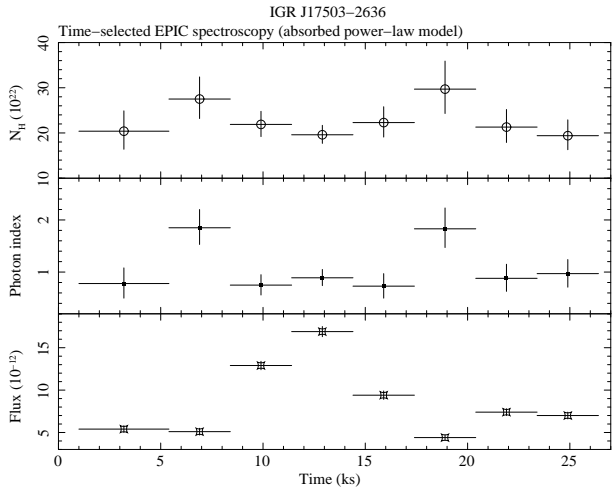} 
 \caption{Time-resolved spectroscopy of \srcon\ (Obs ID 0844101101).
   Eight temporal segments with an exposure time of about 3 ks are reported  (Table~\ref{tab:ontimesel}). 
   In the upper panel is shown the absorbing column density vs time,
   in the middle panel the power law photon index behavior,
    and in the lower panel the average observed flux inside each segment.
}
    \label{fig:ontimesel}
\end{figure}

The light curves of both sources, extracted in two energy ranges (above and below 5 keV)
are shown in Fig.~\ref{fig:lc}, together with the hardness ratios.
Some variability appears in the  hardness of \srcon\ emission
(lower panel on the left of Fig.~\ref{fig:lc}), uncorrelated with
the source count rate. Therefore, we also performed time-selected
spectroscopy of \srcon, dividing EPIC exposure into eight intervals, each about 3 ks in duration.
The best-fit results adopting an absorbed power law are reported
in Table~\ref{tab:ontimesel} and shown in Fig.~\ref{fig:ontimesel}.

\begin{table*}
        \centering
        \caption{Time-selected EPIC spectroscopy of source \srcon,
          adopting a single absorbed power law model (Obs ID 0844101101). The flux is the average observed flux (1-10 keV) in each segment.}
        \label{tab:ontimesel}
        \begin{tabular}{llccrr}
                \hline
no.  & Elapsed Time &   \nh\                  &   $\Gamma$        & Flux             & $\chi^{2}$ (dof)   \\
         & (ks)        &   (10$^{22}$~cm$^{-2}$)   &           &   (10$^{-12}$~\flux) &    \\
\hline
1   & $<$5.3 &    20.4$^{+4.5} _{-4.0}$        & 0.78$^{+0.30} _{-0.28}$ &  5.4$\pm{0.4}$ &   87.15 (85)      \\         
2   & 5.3-8.5 &    27.5$^{+4.9} _{-4.3}$        & 1.85$^{+0.35} _{-0.32}$ &  5.1$\pm{0.4}$ &   75.24 (74)      \\     
3   & 8.5-11.5 &    21.9$^{+2.9} _{-2.7}$        & 0.75$^{+0.20} _{-0.19}$ & 12.9$\pm{0.5}$ &  147.07 (160)      \\
4   & 11.5-14.5 &    19.6$^{+2.1} _{-1.9}$        & 0.89$^{+0.16} _{-0.15}$ & 16.9$\pm{0.6}$  &  206.59 (218)      \\  
5   & 14.5-17.5 &    22.3$^{+3.5} _{-3.2}$        & 0.73$^{+0.24} _{-0.23}$ & 9.4$\pm{0.5}$ &  100.93 (115)      \\         
6   & 17.5-20.5 &    29.7$^{+6.2} _{-5.4}$        & 1.83$^{+0.40} _{-0.36}$ & 4.4$\pm{0.3}$   &   58.43 (66)      \\   
7   & 20.5-23.5 &    21.3$^{+3.9} _{-3.4}$        & 0.88$^{+0.27} _{-0.25}$ & 7.4$\pm{0.4}$  &   83.19 (100)      \\                
8   & $>$23.5 &    19.4$^{+3.5} _{-3.1}$        & 0.97$^{+0.27} _{-0.26}$ & 7.0$\pm{0.4}$  &  116.86 (94)      \\         
\hline
        \end{tabular}
\end{table*}

These eight time-selected spectra are featureless. Steeper spectra appear to be more absorbed. 
However, this trend is led by spectra extracted from 
time intervals 2 and 6. We refit them, fixing the photon index to 0.8, which was obtained in the other temporal segments. 
The resulting  absorbing column densities become
\nh=15.8 $^{+1.7} _{-1.5}$$\times$10$^{22}$~cm$^{-2}$ (spec no.~2) and \nh=17.1 $^{+2.2} _{-2.0}$$\times$10$^{22}$~cm$^{-2}$ (spec no.~6).
The fits are worse ($\chi^{2}$=108.91 for 75 dof in segment no.~2; 
$\chi^{2}$=86.46 for 67 dof in segment no.~6).

\begin{figure}
 \includegraphics[width=6.0cm, angle=-90]{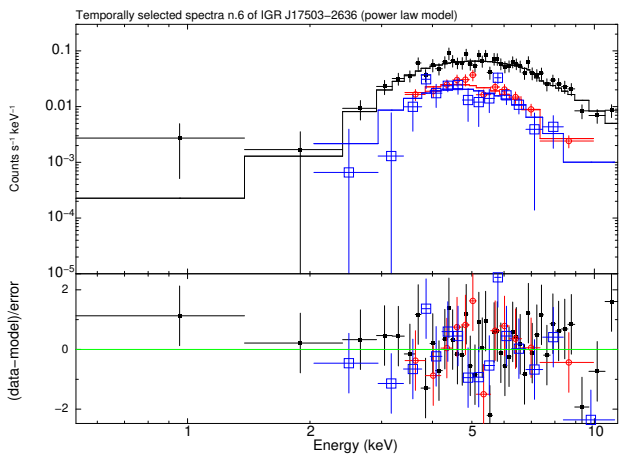} 
 \caption{Spectrum no.6 from the time-resolved spectroscopy of \srcon\ (Obs ID 0844101101).
   The model adopted is a simple absorbed power law.
   Negative residuals appear in both EPIC pn (in black) 
   and MOS2  (in blue). 
   However, this absorption feature is not significant (see text). 
}
    \label{fig:int06spec}
\end{figure}

We note that during the analysis of spectra extracted from time interval no.~6 (Fig.~\ref{fig:int06spec}), 
some negative residuals appear at energies around 9.5 keV. 
We used a cyclotron line model ({\sc cyclabs} in {\sc xspec}, multiplying the absorbed power law)
to account for it.
However, we tested the significance of the line running 10$^{6}$ simulations
with the tool {\sc simftest} in {\sc xspec}. 
It returned a probability of 11\% that data are consistent with the model without an extra component, implying that this feature is not significant.

\subsubsection{\xmm\ ObsID 0932190801}
\label{sect:xmmrestwo}

During this observation, \igrtre\ was undetected, while \igrsette\ was bright enough
for meaningful spectroscopy. We report the results in Table~\ref{tab:specigrsette0801}. 
The EPIC light curve of \igrsette\ was very similar to the one observed in Fig.~\ref{fig:lc}, with no evident variability in both intensity and the hardness ratio, along the whole exposure.

\begin{table}
        \centering
        \caption{Time-averaged EPIC spectroscopy of  \igrsette\ (Obs ID 0932190801).
          The flux (UF) has been corrected for absorption (1-10 keV). 
        d$_{10\,kpc}$ is the source distance in units of 10 kpc.  }
        \label{tab:specigrsette0801}
        \begin{tabular}{lcc}
                \hline
        Param                               &                              &        \\
                \hline
   \nh\  (10$^{22}$~cm$^{-2}$)              &     7.4$^{+2.2} _{-1.8}$     &    7.5$^{+2.3} _{-1.9}$   \\   
Power law $\Gamma$                          &     0.28$^{+0.30} _{-0.28}$  &     0.35$^{+0.33} _{-0.30}$  \\
UF (10$^{-11}$~\flux)                       &   0.45$^{+0.04} _{-0.04}$    &   0.44$^{+0.04} _{-0.04}$    \\
E$_{line}$ (keV)                            & $-$                          &   6.31$^{+0.10} _{-0.10}$  \\
line width  (keV)                           &         $-$                  &    0.0 (fixed)       \\
line norm ($10^{-6}$ ph cm$^{-2}$ s$^{-1}$) &         $-$                  &    13$^{+7} _{-7}$     \\
EW (eV)                                     &         $-$                  &   250$\pm{130}$            \\
L$_{\rm X}$ (10$^{34}$ $d^2_{10\,kpc}$ erg~s$^{-1}$)    &     5.4          &    5.3  \\
$\chi^{2}$ (dof)                            &      72.75 (55)              &    62.39 (53)     \\
                \hline
        \end{tabular}
\end{table}

We calculated upper limits for the source flux, adopting the {\sc eupper} tool of the {\sc sas} on MOS2 data,
which was the only camera for which the source position was not 
too close to the chip gaps.
The local background  was evaluated from a circular region near the source, at a similar off-axis position ($\sim$ 12.3 arcmin), where the effective exposure time reduces to about 3.2 ks.
In Table~\ref{tab:xmmeupper}, we list the 3$\sigma$ upper limits to the X-ray flux of  \igrtre. 
The flux corrected for absorption (UF in this table), calculated in the energy range 2-10 keV, is also reported in the last column, assuming a power law model with a photon index of $\Gamma$=1 and an absorbing column density of N$_{H}$=2$\times$10$^{23}$\,cm$^{-2}$.

\subsubsection{\xmm\ ObsID 0932191001}
\label{sect:xmmresthree}

During this observation of the Milky Way survey, only the sky position
of \igrtre\ was inside the FoV, but the source is undetected.
In order to evaluate upper limits on the source flux, we considered EPIC pn and MOS1 products, but not MOS2, which is strongly contaminated by stray light at the source position.
The source is located at an off-axis position of $\sim$12.5 arcmin, where the exposure time reduces to about 2.1 ks and 3.3 ks for the pn and MOS~1, respectively.
The  3$\sigma$ upper limits are reported in Table~\ref{tab:xmmeupper}.
As before, the flux, corrected for absorption (UF; 2-10 keV), was calculated assuming a power law model with a photon index of $\Gamma$=1 and an absorbing column density of N$_{H}$=2$\times$10$^{23}$\,cm$^{-2}$.

\begin{table}
  \caption[]{\xmm\ 3$\sigma$  upper limits on the X-ray emission from IGR\,J17503-2636. UF is the flux, corrected for absorption, evaluated assuming   a power law model with $\Gamma$=1 and an absorbing column density of N$_{H}$=2$\times$10$^{23}$\,cm$^{-2}$.
}
\begin{tabular}{lllll}
\hline
\noalign {\smallskip}
  ObsID        & Instr.        & Count rate (2-12 keV)   &  UF (2-10 keV)   \\
               &                  &  (counts s$^{-1}$)      &    \flux\  \\
  \hline
  0932190801   &   MOS 2        &    $<$0.0077            &   $<$6.3$\times10^{-13}$   \\  
  0932191001   &   pn           &    $<$0.0042            &   $<$9.5$\times10^{-14}$    \\  
  0932191001   &   MOS 1        &    $<$0.0035            &   $<$2.8$\times10^{-13}$    \\  
\hline
\label{tab:xmmeupper}
\end{tabular}
\end{table}

\subsubsection{Timing analysis}
\label{sect:timing}

\begin{figure}
\centering
\includegraphics[width=8.0cm]{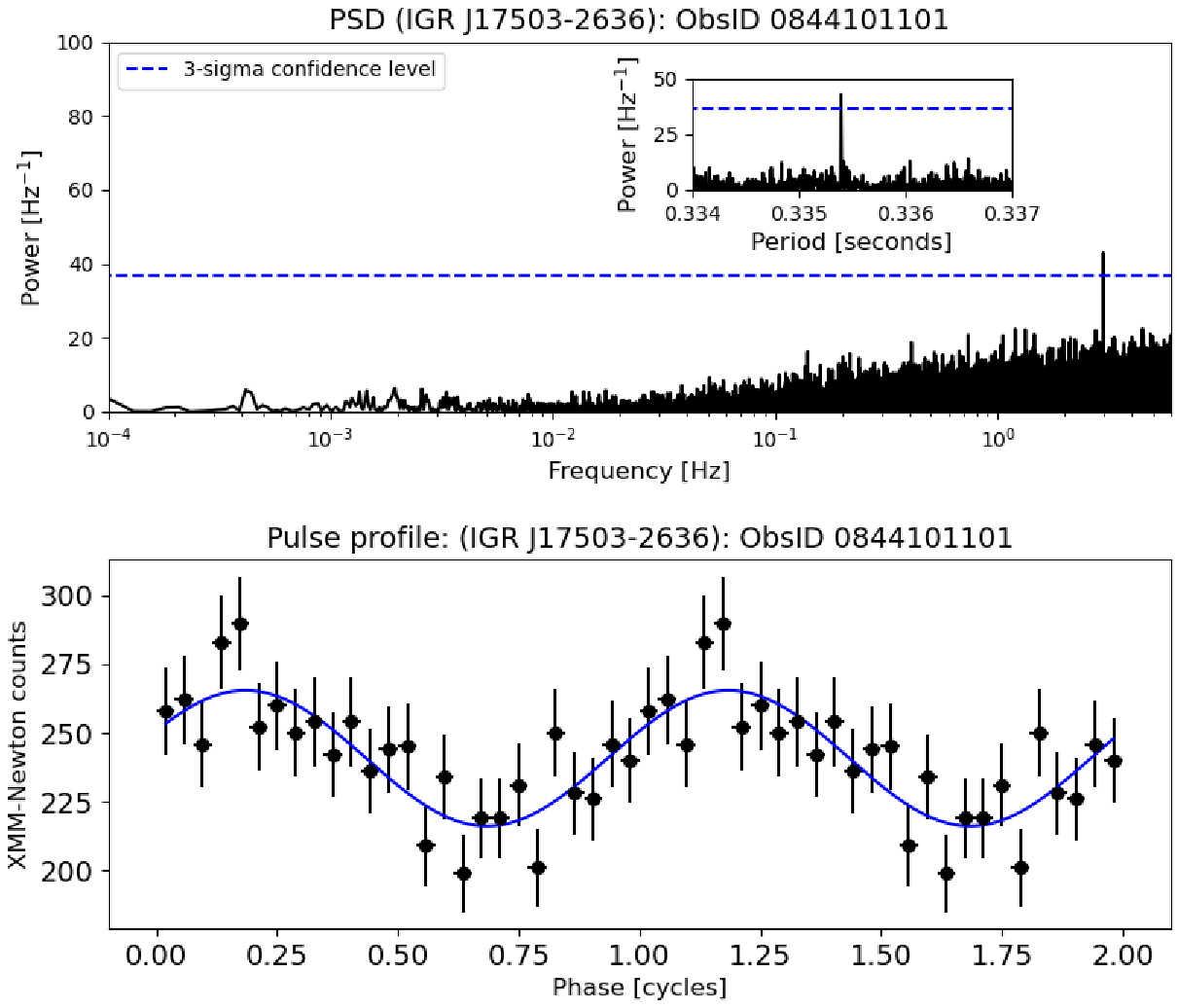}
\caption{PSD of the EPIC-pn observation of IGR J17503-2636 (ObsID 0844101101) with the 3$\sigma$ confidence level plotted. Bottom panel: Pulse profile (and best-fit sinusoid) corresponding to the periodicity candidate with a period of 0.3 seconds.}
\label{Fig:timing}
\end{figure}

We performed a search for periodic pulsations in the two {\it{XMM-Newton}} EPIC-pn observations (ObsIDs: 0844101101 and 0932190801) for the two sources IGR J17503-2636 and IGR J17507-2647, by implementing a Fourier domain acceleration search method via the {\tt{PRESTO}} software (see \citealt{Ransom2002} for more details on the method). The event files were barycentered using the {\tt{barycen}} tool with the source positions derived from the EPIC-pn observations. We then extracted events in the 4-8 keV energy range from a circular region of 15 arcsec radius centered on the source position (for IGR J17503-2636) and an elliptical region (for IGR J17507-2647), created light curves binned to the optimal EPIC-pn resolution of 73 ms, and subtracted the local background. Using this background subtracted light-curve and the {\tt{accelsearch}} routine in {\tt{PRESTO}}, we detected a possible periodicity candidate for IGR J17503-2636, as is shown in the power spectral density (PSD) in Fig.~\ref{Fig:timing}. 
This candidate was found to have a period of 0.335397(3) seconds, period derivative, \pdot, of ($5.1^{+0.5}_{-7})\times 10^{-10}$ s~s$^{-1}$,
and a pulsed fraction of (10 $\pm$ 1)\%, with a significance of $\sim 3.8\sigma$ (after considering the total number of trials in the PSD). In order to correct for the low-frequency (red) noise, we divided the observed PSD by the best-fit phenomenological (constant + power-law) model for the red noise. Moreover, in order to verify whether the periodicity could be associated with the background, we extracted events of the background in the vicinity of the source and ran the same search as above. We did not find any periodicity candidates associated with the background above a significance of 1.5$\sigma$.

As a caveat, we note that the significance of the periodicity for IGR J17503-2636 peaks in the 4-8 keV energy range, and has a lower significance of 2$\sigma$ if events between 2-4 keV are also included. The significance also appears to vary with the size of the source extraction region (with a significance of 3.8$\sigma$ for a radius of 15 arcsec and a lower significance of 2$\sigma$ for a radius of 20 arcsec). 
As a final remark, we note that we could not find any additional archival X-ray dataset to check this candidate periodicity, including Jem-X data during the discovery outburst, where the source could only be detected \citep{Chenevez2018, Ferrigno2019}.

In the case of IGR J17507-2647, we did not detect any periodic signals above a significance level of 2$\sigma$ (in the range of 0.2-1000 s). We placed an upper limit on the pulsed fraction of less than $20$\% (for the candidate with the highest power in the PSD with period of $\sim$ 37 seconds).

\subsection{\sw}
\label{sect:swiftres}

\begin{figure*}
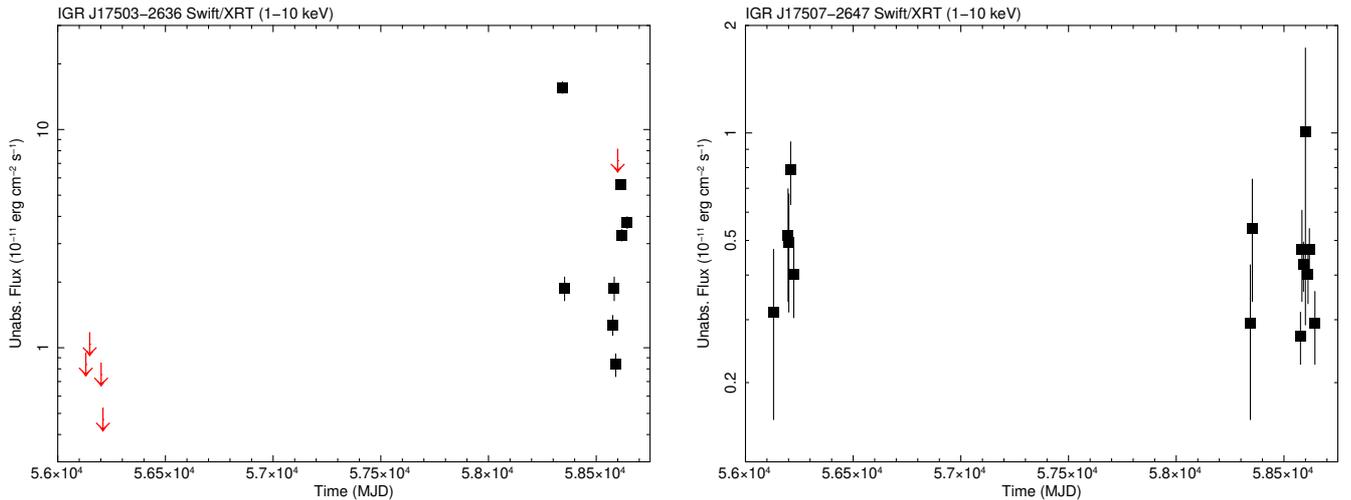

\begin{center} 
\resizebox{\hsize}{!}{\includegraphics[width=6.0cm,angle=-90]{igr17503_swift_uf_new.ps}
 \includegraphics[width=6cm,angle=-90]{igr17507_swift_uf_new.ps}}
  \caption{\sw/XRT long-term light curves with fluxes corrected for absorption in the 1-10 keV energy range (\igrtre\ on the left, \igrsette\ on the right). Time is in units of MJD (spanning seven years, from July 2012 to May 2019). Fluxes are in units of 10$^{-11}$~\flux. 
  Downward red arrows indicate 3$\sigma$ upper limits on the source intensity.}
\label{fig:swflux}
\end{center}
\end{figure*}

For completeness, we have analyzed the 
\sw/XRT observations covering the sky positions of both sources, to obtain their long-term light curves. 
However, we note that most of the \sw/XRT observations of \igrtre\ have already been reported and discussed by \citet{Ferrigno2019} and \citet{Ferrigno2022}.
Net count rates (0.3-10 keV; per-observation) are reported in Table~\ref{tab:swrates}, while the light curves are shown in Fig.~\ref{fig:swflux}, with fluxes (1-10 keV) corrected for absorption. The conversion from net count rates to unabsorbed fluxes assumes a power law model with a photon index, $\Gamma$, of 1 and a column density of \nh=2$\times10^{23}$~cm$^{-2}$ for \igrtre\ and \nh=1$\times10^{23}$~cm$^{-2}$ for \igrsette.
The \sw\ long-term light curve of \igrsette\ shows a stable intensity, with 
an average flux, corrected for absorption, of 4$\times10^{-12}$~\flux\ (1-10 keV).

\begin{table}
        \centering
        \caption{\sw/XRT count rates (0.3-10 keV) for both sources. Upper limits are at 3$\sigma$.}
        \label{tab:swrates}
        \begin{tabular}{lcc}
                \hline
ObsID    &   \igrtre                    &   \igrsette  \\
         &  (10$^{-2}$~count~s$^{-1}$)  &  (10$^{-2}$~count~s$^{-1}$)    \\
\hline
10807001$^{a,b}$  &    46.6$\pm{2.9}$         & 1.3$\pm{0.6}$  \\
10980001$^{b}$    &    3.8$\pm{0.4}$          & 1.2$\pm{0.2}$  \\
10980002$^{b}$    &    5.6$\pm{0.7}$          & 2.1$\pm{0.6}$  \\
10980003$^{b}$    &    2.5$\pm{0.3}$          & 1.9$\pm{0.3}$  \\
10980004          &    $<$21.5                & 4.5$\pm{3.2}$  \\
10980005$^{b}$    &    17.0$\pm{0.9}$         &  1.8$\pm{0.3}$ \\    
10980006$^{b}$    &    9.8$\pm{0.6}$          & 2.1$\pm{0.3}$  \\
10980007$^{b}$    &    11.2$\pm{0.7}$         & 1.3$\pm{0.3}$  \\
43694001          &    $<$2.5                 & 1.4$\pm{0.7}$  \\
43701001          &    $<$3.1                 &    $-$  \\
48022002          &    $-$                    & 2.3$\pm{0.8}$  \\    
48022003          &    $<$2.25                & 2.2$\pm{0.8}$  \\    
48022004          &    $<$1.4                 & 3.5$\pm{0.7}$  \\
48022007$^{b}$    &   stray light cont.       &  1.8$\pm{0.4}$      \\
88805001$^{b}$    &    5.6$\pm{0.7}$          &  2.4$\pm{0.9}$    \\
\hline
        \end{tabular}
\footnotesize{\\
$^a$ Observation reported also by \citealt{Ferrigno2019}  about \igrtre; $^b$ Observation reported also by \citealt{Ferrigno2022} about \igrtre. \\
}
\end{table}

\subsection{\inte}
\label{sect:interes}

Both sources were in the FoV of IBIS on board the INTEGRAL satellite. We discuss these sources separately in the following subsections.

\subsubsection{\igrtre}
\label{sect:interestre}

\igrtre\ is a transient source discovered by INTEGRAL  on August 11, 2018, during satellite revolution 1,986 \citep{Chenevez2018, Ferrigno2019}.   
We utilized the bursticity method developed by  \citealt{Bird2010, Bird2016} 
in order to search for any possible transient hard X-ray  activity from the source before its discovery. To this aim, we used all public IBIS/ISGRI observations during which the source was in the instrument FoV; that is, from revolution 46 (February 28, 2003) to 1,945 (April 23, 2018).  No significant transient  hard X-ray activity was found in the  energy band 18-60 keV.
\igrtre\  is not reported in the latest published INTEGRAL IBIS catalog (up to revolution 1,000) of \citealt{Bird2016},
despite an extensive coverage of its sky  region for an effective   exposure time of $\sim$12 Ms. This information can be used to infer a 3$\sigma$ upper limit  of 0.3 mCrab or 2.3$\times$10$^{-12}$ erg cm$^{-2}$ s$^{-1}$ (20--40 keV) on its hard X-ray flux in quiescence.

\citealt{Ferrigno2019} report a duration of $\sim$12 days for the transient X-ray activity of  IGR J17503$-$2636. 
However, we note that the true duration could  be significantly longer: in fact, the source was discovered  during INTEGRAL satellite revolution 1,986 (August 2018), and it was not in the IBIS/ISGRI FoV during any previous  satellite revolutions back to revolution 1,946 (April 23, 2018). 
Because of this temporal gap of $\sim$4 months, the duration of the source outburst activity  is unconstrained. In principle,  it could well be that \igrtre\ was already active when it entered for the first time in the satellite  FoV when discovered. This hypothesis is supported i) by the relative weakness of the source 20-40 keV flux ($<$8$\times$10$^{-11}$ erg cm$^{-2}$ s$^{-1}$)  when discovered, ii) by the evident declining trend of the source flux when discovered, as has clearly been measured in the soft X-ray band by different X-ray satellites \citep{Ferrigno2019}. 
All this suggests that the brightest source peak activity had already happened  when it entered for the first time in the INTEGRAL FoV and was discovered.
Since the source was not detected during the satellite revolution 1,945 (April 23, 2018) -- the closest observation  during which  the source was again in the INTEGRAL  FoV before its discovery in revolution 1,986 -- we can place an upper limit of $\sim$4 months on the outburst duration.

\subsubsection{\igrsette}
\label{sect:interesette}

\igrsette\ is a persistent hard X-ray source best detected in the energy band 18-60 keV. 
It is characterized by a very stable flux of 1.1 mCrab or 8.3$\times$10$^{-12}$ erg cm$^{-2}$ s$^{-1}$  
(20-40 keV), as is consistently listed in  all the  published INTEGRAL  catalogs \citep{Bird2007, Bird2010, Bird2016}.
We produced and investigated the long term IBIS/ISGRI light curve (18-60 keV), spanning a baseline of $\sim$7 years of observations, in order to highlight any possible flaring or strongly variable behavior of the source on top of its persistent nature, with negative results.

\section{Discussion}
\label{sec:discussion}

We have analyzed  unpublished X-ray observations of two IGR sources whose nature
is uncertain. 
The  \xmm\ observations have permitted us to
obtain the most sensitive view in soft X-rays of \igrtre\ and \igrsette to date:
we have constrained the lowest luminosity state ever observed from \igrtre\ and
discovered a broad iron emission line  in the \igrsette\ spectrum, with centroid energy in the range of 6.3-6.6 keV.
In parallel, we have analyzed INTEGRAL data, spanning a baseline of about 15 years.
We discuss these results in the following subsections.

\subsection{\igrtre}

The \xmm\ observations confirm the X-ray transient behavior of this
source, allowing us to pose the most stringent 3$\sigma$ upper limit on the source flux
UF$<$9.5$\times10^{-14}$\flux\ (2-10 keV, corrected for absorption).
This is the lowest X-ray emission constrained for this source, leading to a dynamic range larger than $\sim$2100, compared with the flux in outbursts measured by Jem-X (2$\times10^{-10}$~\flux; \citealt{Chenevez2018}). 
Our upper limit  translates into an X-ray luminosity
L$_X$$<$1.1$\times$10$^{33} d^2_{10\,kpc}$~erg~s$^{-1}$ (2-10 keV), where d$_{10\,kpc}$ is the source distance in units of 10 kpc.
 
During the pointed \xmm\ observation performed in 2020,
the source displayed a very variable X-ray light curve, with some flaring behavior, typical of a wind-fed XRB \citep{Martinez2017}.
The time-average flux was 1.7$\times10^{-11}$~\flux\ (1-10 keV),
showing a highly absorbed spectrum (\nh=2$\times10^{23}$~cm$^{-2}$), 
well described by a featureless hard power law model ($\Gamma$=1). 
Temporally resolved spectroscopy showed a variable power law slope, with a steeper power law when the spectrum was more absorbed.

Our INTEGRAL results highlight the transient nature of IGR J17503$-$2636. From satellite observations covering a baseline of  
$\sim$15 years (2003-2018), the source has been detected in an outburst only once, when it was discovered.  It  spends the great majority of the time undetected with an hard X-ray flux lower than 0.3 mCrab or 2.3$\times$10$^{-12}$ erg cm$^{-2}$ s$^{-1}$ (20-40 keV).  At odds with previous literature reporting on short, hard X-ray activity (i.e., \citealt{Ferrigno2019}), we have demonstrated that the duration of the discovery outburst cannot be exactly determined. However, it cannot be longer than $\sim$4 months. 

\citet{Masetti2018} identified the NIR counterpart of \igrtre, and interpreted the
NIR photometry with emission from a highly absorbed early type supergiant companion,
proposing an SFXT nature.
SFXTs are HMXBs where a neutron star (NS) accretes matter from the wind of 
an O- or B-type supergiant companion \citep{Sguera2005, Sguera2006, Negueruela2006, Kretschmar2019}.
SFXTs show brief outbursts (lasting only a few days),
punctuated by short flares during which the X-ray luminosity reaches
10$^{36-37}$ erg~s$^{-1}$ for a few thousand seconds  (with a duty cycle of less than 5\%; \citet{Sidoli2018}).
SFXTs spend the majority of their lifetime below 10$^{34}$ erg~s$^{-1}$, down to quiescence, when the 
luminosity is around 10$^{31-32}$~erg~s$^{-1}$ \citep{Kretschmar2019, Sidoli2023}.

\citet{Ferrigno2019, Ferrigno2022} discuss  the X-ray light curves and spectra,
supporting the proposed SFXT identification of \igrtre, although
they also note something anomalous for an SFXT: the long X-ray outburst and the limited range of X-ray flux
variability of a factor of 300 (compared with the typical 10$^{3}$-10$^{4}$).
Incidentally, Ferrigno et al. mention an alternative interpretation of
the NIR counterpart \citep{McCollum2018} in a note, but they do not discuss it further  and
take into consideration only the SFXT scenario.

On the contrary, we think that the results reported by McCollum et al. (2018) deserve attention, since they represent a significant step forward
in the analysis of the counterpart: first of all, their dataset extends over a broader range of wavelengths, up to the MIR range, considering also {\em Spitzer} and WISE  observations. 
Moreover, these authors performed a fit of the SED, finding that the best fit is an M-type giant star (M4-6 III star), with an  excess at longer wavelengths than 12 $\mu$m. It is remarkable that a hot massive star resulted in a significantly  worse fit.
Therefore, in the absence of confirmation by means of IR spectroscopy,
we cannot ignore this published result, implying that
\igrtre\ is not an SFXT, but a different kind of XRB.

X-ray binaries with M-type giants companions are called symbiotic X-ray binaries (SyXBs; \citealt{Masetti2006}, \citealt{Yungelson2019}).
SyXBs are a small (13 members, according to \citet{Yungelson2019})
subclass of wide binaries (orbital periods $>$100 days) 
composed of an NS accreting from the wind of a late-type (K1-M8) giant star. 
They display transient X-ray activity, with X-ray luminosity variable 
in the range of 10$^{32}$-10$^{36}$ erg s$^{-1}$. 
Some members show  X-ray pulsations with long periodicities (from hundred to thousands of seconds).
Models to explain their X-ray outbursts share similarities to those proposed for SFXTs. 
This is not surprising, since these two types of highly magnetized NS wind accretors
share many similarities in their observed X-ray properties 
(e.g., their flux variability and flaring behavior), 
the donor star (late vs. early type (super)giant star) and their outflowing stellar wind
being the main crucial difference.
The proposed models include the propeller regime \citep{Afonina2024}, and the
quasi-spherical settling accretion model. This latter was originally developed for HMXB pulsars \citep{Shakura2012}, 
later applied to SFXTs \citep{Shakura2014}, and then adapted to SyXBs  \citep{Yungelson2019}. 
 
In order to apply them to the single sources, the NS magnetic field strength and the pulsar spin period are crucial quantities.
%
For what concerns the NS magnetic field, a hint of a  cyclotron scattering feature 
was reported by \citep{Ferrigno2019} from the analysis of the \nustar\ spectrum of \igrtre. 
If confirmed by future observations, this implies an NS magnetic field of B$\sim$2$\times10^{12}$~Gauss.
Our timing analysis of \xmm\ data led to a barely detected periodicity at $\sim$0.3 s.
Since this periodicity needs to be confirmed, the discussion about the nature of \igrtre\ can only be speculative, also in light of the fact that an NS with a spin period of 0.3~s 
in a binary system with an M giant star has no analogs, to the best of our knowledge. 
Searching the most recent catalog of LMXBs (including the SyXBs)\footnote{http://astro.uni-tuebingen.de/~xrbcat/LMXBcat.html}  \citep{Avakyan2023}, we found 
only the sources GRO\,J1744-28 (P$_{spin}$=0.467 s)  
and 3A\,1822-371 (P$_{spin}$=0.59 s) with similar pulse periods. 
While for this latter source the spectral type of the donor star is not reported, the companion of the bursting pulsar GRO\,J1744-28 is a G/KIII star \citep{Gosling2007, Doroshenko2020}.

If the candidate pulse period is confirmed, and if the system is fed by the wind of the giant companion, the spherical settling accretion model can be ruled out, as it applies to much slower pulsars \citep{Shakura2012, Yungelson2019}. 
In this case, a wind-fed system (with a pulse period of less than $\sim$30 s) is in the Bondi-Hoyle accretion stage.

Yungelson et al. (2019)  presented their model of population of SyXBs, predicting the distributions of sources in the NS spin period–X-ray luminosity plane (P$_{spin}$–L$_{\rm X}$; their Fig.\,2) for  optical companions at different evolutionary stages.
Remarkably, \igrtre\ would be the first SyXB located in a predicted high number density region of the P$_{spin}$–L$_{\rm X}$ diagram, which is still not populated by any known SyXBs (i.e. Log(P$_{spin}$)=-0.5 and L$_{\rm X}$$\sim$10$^{35}$).

The fact that \igrtre\ is undetected during two \xmm\ observations reported here might be explained by the onset of the propeller (the centrifugal barrier halts most of the accretion). 
If this is the case, the limiting accretion luminosity for the onset of the propeller (eq.~1 in \citealt{Campana2002}) is

\begin{eqnarray}
L_{\rm lim}&\simeq& 3.9\times 10^{37}
\,\xi^{7/2}\,B_{12}^2\,P_{spin}^{-7/3}\,M_{1.4}^{-2/3}\,R_6^{5}\ergs 
,\end{eqnarray}
where the spin period is in units of seconds, the NS magnetic field, $B_{12}$, is in units of $10^{12}$ G, and the NS mass and radius 
scale as $M=M_{1.4}\,1.4\msun$ and $R=R_6\,10^6$ cm. 
We assume the constant factor $\xi=1$ for a spherical 
symmetry, and $M_{1.4}$=1 and $R_6$=1 for the parameters of the NS.
If P$_{spin}$=0.3~s and B$_{12}$=2, \igrtre\ should always be in a
propeller state, 
as L$_{\rm lim}$=2.6$\times$10$^{39}$~erg~s$^{-1}$, which is
orders of magnitude larger than the accretion X-ray luminosity observed with \xmm\ (for a source in our Galaxy).
In conclusion, both candidate values for these NS quantities cannot be true at the same time in \igrtre: if one of them is confirmed, either B is much lower or the NS rotates much more slowly.
Finally, we remark that this conclusion is not affected by the assumptions made about the parameters on the right hand side of Eq. 1. It still holds even for an NS with a 2.5$\msun$ mass and a radius of 12 km, or for $\xi=0.5$. The limiting X-ray luminosity from Eq. 1 will always be much larger than the X-ray luminosity of the source, even in the case of a very large distance  (i.e., 20 kpc).

\begin{figure}
  \includegraphics[width=9.0cm, angle=0]{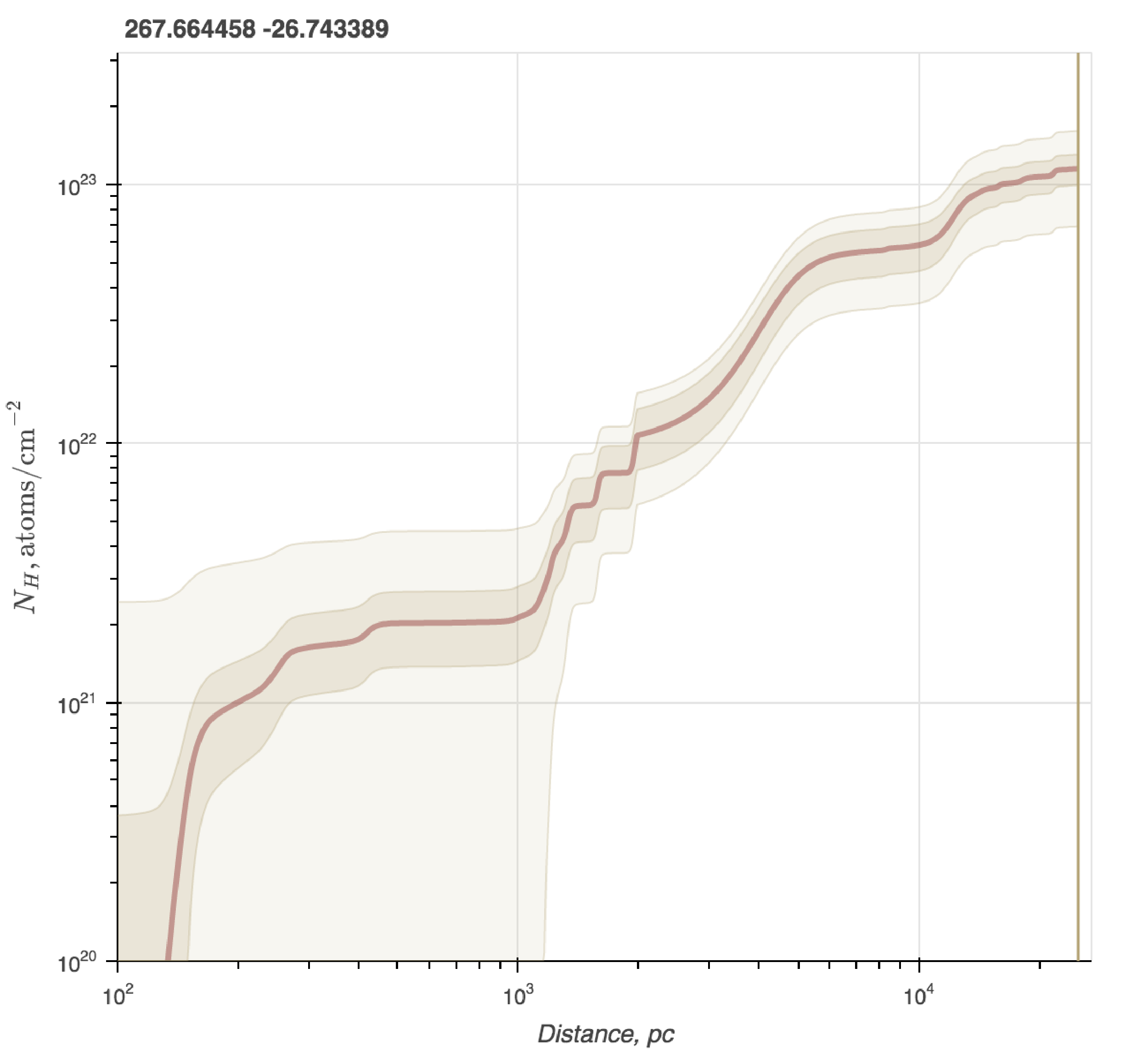}
  \caption{Absorption vs. distance in the direction of \igrsette,
    obtained using the online tool  
    {\sc http://astro.uni-tuebingen.de/nh3d/nhtool} \citep{Doroshenko2024}.
    The shaded regions show the estimated uncertainties
    for the estimate of N$_{H, X-ray}$ (light color) and N$_{H, E(B-V)}$ (denser color).
}
    \label{fig:nh}
\end{figure}

\subsection{\igrsette}

Besides the INTEGRAL catalog \citep{Bird2007}, the only papers in the literature
discussing this source are \citet{Tomsick2009}, who identified its \chandra\ counterpart, and
\citet{Zolotukhin2011}, who found a faint, barely detected NIR counterpart in the UKIDSS-DR3 Galactic Plane Survey (K=18.5$\pm{0.4}$~mag, J$>$20.3~mag, H$>$19.3~mag).

We have analyzed two unpublished \xmm\ observations covering the source sky position. In both datasets, the source is detected
at a similar flux, corrected for absorption, of 4$\times$10$^{-12}$~\flux (1-10 keV).
This is also consistent with the \chandra\ flux \citep{Tomsick2009},
and with the long-term light curve we have obtained from \sw/XRT observations,
indicative of a remarkably stable, persistent X-ray emission.
The EPIC spectrum is highly absorbed (\nh$\sim$10$^{23}$~cm$^{-2}$) and well fit by a flat power law model (photon index in the range from 0 to 1), which is again consistent with the \chandra\ results.
With respect to the \chandra/ACIS-I spectrum, we have detected an iron emission line (at 6.3-6.6 keV) in both \xmm\ observations, mostly consistent with a Fe K$_{\alpha}$ line.

We have investigated the long-term INTEGRAL IBIS light curve (18-60 keV) spanning a baseline of $\sim$7 years of observations
in order to search for any possible flaring or strongly variable behavior of the source on top of its persistent nature.  We found that 
 \igrsette\  is characterized by a very stable, persistent, and weak hard 
X-ray emission at a level of $\sim$8$\times$10$^{-12}$ erg cm$^{-2}$ s$^{-1}$ (18-60 keV).  This is a behavior  similar 
to that found in the softer X-ray band (1-10 keV).  

In previous literature, the source nature was discussed by \citet{Tomsick2009}.
These authors, based on the lack of a 2MASS, DENIS, USNO-B1.0, or USNO-A2.0 counterpart, suggested that the high absorption derived from the ACIS-I spectrum may be of interstellar origin, implying a large distance (i.e., close to the Galactic center). In this case, they derived an X-ray luminosity of  $\sim$4$\times$10$^{34}$~erg~s$^{-1}$ (at 8.5 kpc). 
\citet{Tomsick2009} concluded that this luminosity is too large for a CV, favoring an XRB.
On the other hand, an LMXB nature was simply assumed by \citet{Zolotukhin2011},  
because of its location in the direction of the Galactic bulge region.
Among XRBs, we note that the flat power law spectrum observed both during \chandra\ and \xmm\ observations
disfavors a LMXB, being more typical of an accreting pulsar in a HMXB, or a CV. 
But since a HMXB can be ruled out based on the candidate NIR counterpart (see below), we believe that the CV nature is a viable possibility.

We have used the online tool 3D-N$_{\rm H}$\footnote{http://astro.uni-tuebingen.de/nh3d/nhtool} 
\citep{Doroshenko2024}
to obtain the dependence of the absorbing column density, \nh\ , from the distance toward \igrsette\ (we have downloaded this plot and reported it in Fig.~\ref{fig:nh} with permission by V. Doroshenko).
Assuming that the N$_H$ we measured (\nh\ in the range of 5-10$\times$10$^{22}$ cm$^{-2}$) in EPIC spectra is purely interstellar (as was also suggested by \citet{Tomsick2009}), this implies a range of values for the pair of quantities, A$_V$, and source distance (d) in between the two extremes, 
(1) A$_V$$\sim$15 mag and d$\sim$6 kpc (for \nh=5.$\times$10$^{22}$ cm$^{-2}$) and
(2) A$_V$$\sim$29 mag and d$\sim$16 kpc (for \nh=1.$\times$10$^{23}$ cm$^{-2}$).

In these two cases, the NIR counterpart excludes a hot massive star in our Galaxy, even on the main sequence (e.g., B0V), ruling out any type of HMXB.
Therefore, the flat X-ray spectra below 10 keV leaves room only for a CV. 
K-M main sequence stars might match the NIR counterpart of \igrsette. This would suggest a CV or an LMXB. However, in these cases a significant contribution from an accretion disk might be present \citep{demartino2020}, so we think that it is premature to engage in any further speculation about a single K-filter measurement. 
Moreover, we disfavor a LMXB, because these X-ray sources typically display a much steeper X--ray
spectrum ($\Gamma$$\sim$2) below 10 keV.
If we consider a source distance of $\sim$6 kpc, resulting in an X-ray luminosity of 2$\times$10$^{34}$ erg s$^{-1}$, this is typical of magnetic CVs (mCVs), hosting a white dwarf with a strong magnetic fields (B$>$10$^6$ Gauss). 
The mCVs are observed in a wide range of X-ray luminosities of $10^{30}$--$10^{35}$ erg s$^{-1}$ and a few times $10^{34}$ erg s$^{-1}$ is a typical value in intermediate polars (IPs; \citealt{demartino2020, Suleimanov2022}). 
The spectra of IPs are characterized by flat power-law shape with $\Gamma\sim1$ and iron complex emission at 6-7 keV (a 6.4 keV fluorescent iron line, an Fe XXVI line at 6.9 keV, and an Fe XXV line at 6.7 keV; \citet{demartino2020}). 
The mean equivalent width (EW) of the ionized 6.7 keV line in IPs is around $107\pm{17}$ eV \citep{xu2016}. 
The Fe K$\alpha$ at 6.4 keV detected by \xmm\ , together with the upper limits we placed on the equivalent width of the ionized iron lines, are also consistent with a mCV nature. 
The CV nature is also in line with our INTEGRAL results on \igrsette: if located at 6 kpc, the hard X-ray luminosity is $\sim$3.5$\times$10$^{34}$~erg~s$^{-1}$ (18-60 keV). In this context, we note that to date all the CVs  detected by INTEGRAL above 20 keV are stable, weak, and persistent hard X-ray  sources with typical luminosities in the range of 10$^{32-34}$~erg~s$^{-1}$ \citep{Barlow2006, Revnivtsev2008, Brunschweiger2009}.

Therefore, we conclude that an mCV nature is favored for \igrsette.

 \section{Conclusions}
 \label{sect:concl}

We have reported and discussed X-ray results (\xmm, \sw\  , and \inte) for 
two hard X--ray sources,  \igrtre\ and \igrsette.
The main new results on \igrtre\ can be summarized as follows:

\begin{itemize}
    \item The analysis of an \xmm\ observation constrains the source flux in the lowest intensity state ever recorded for this source ($<$9.5$\times$10$^{-14}$~\flux; 3$\sigma$, 2-10 keV,  corrected for absorption), updating the amplitude of flux variability (2-10 keV), compared with previous observations.
    \item The \xmm\ light curve  is highly variable, with several short ($\sim$1~ks) flares, typical of wind-fed X-ray sources.
    \item A candidate pulsation at 0.3 s is barely detected (at 3.8$\sigma$) in EPIC data, which needs confirmation.
    \item If this candidate spin period is confirmed by further observations, it cannot coexist with a high NS magnetic field of 2$\times$10$^{12}$~G (reported previously in the literature as a candidate value, which needs confirmation too); otherwise, the source would always be in a propeller (i.e., mostly not accreting) state.
    \item The \xmm\ spectrum is very absorbed (\nh=2$\times$10$^{23}$~cm$^{-2}$), featureless, and flat (power law photon index of $\Gamma$$\sim$1, with evidence of  variability along the 25~ks EPIC exposure).
    \item The duration of the INTEGRAL discovery outburst (the only one bright enough to be detected by \inte) has been revised and can be constrained to be shorter than $\sim$4 months (much longer than the duration of $\sim$12 days previously reported in the literature).
    \item This information, put into the context of the alternative identification of the infrared/optical counterpart of \igrtre\ with an M-type giant \citep{McCollum2018},
    has led us to propose that\igrtre\ is a SyXB, instead of a SFXT.  In fact, we believe that the fit of the SED of the companion star \citep{McCollum2018} over a large range of energies (compared with Masetti et al. photometry) represents a significant step forward in the identification of the source nature, in the absence of IR/optical spectroscopy.
\end{itemize}

The main new results that we have obtained for \igrsette\ can be summarized as follows:
\begin{itemize}
 \item The \xmm\ observations show stable and persistent X-ray emission, with a hard X-ray spectrum and a broad iron emission line with centroid energy at $\sim$6.3-6.6 keV; an iron emission line has never been observed before from this source.
 \item The source was always detected during the short \sw/XRT observations spanning 7 years (July 2012-May 2019), with an average flux of 4$\times$10$^{-12}$~erg~cm$^{-2}$~s$^{-1}$ (1-10 keV, corrected for absorption).
  \item The INTEGRAL data highlight the persistent character of its hard X-ray emission, consistent with the properties of mCVs observed by INTEGRAL.
  \item These new findings have led us to propose that \igrsette\ is a distant mCV. 
\end{itemize}


 \begin{acknowledgements}
This research is based on observations performed by the satellites \xmm,
INTEGRAL and \swift.
This research has made use of data and software provided by
the High Energy Astrophysics Science Archive Research Center (HEASARC), 
which is a service of the Astrophysics Science Division at NASA/GSFC.  
This research has made use of the VizieR catalog access tool, CDS,
Strasbourg, France  \citep{vizier2000}.
We acknowledge the use of public data from the \sw\ data archive.
We made use of the tool 3D-N$_{\rm H}$ ({\sc http://astro.uni-tuebingen.de/nh3d/nhtool}) made publicly available by V.~Doroshenko.
LS, VS and PE acknowledge funding from INAF through the grant ``Bando Ricerca Fondamentale INAF 2023". 
RS and PE acknowledge financial support from the Italian Ministry for University and Research, through the grants 2022Y2T94C (SEAWIND) and from INAF through LG 2023 BLOSSOM. 

\end{acknowledgements}

\bibliographystyle{aa}

\begin{thebibliography}{47}
\expandafter\ifx\csname natexlab\endcsname\relax\def\natexlab#1{#1}\fi

\bibitem[{{Afonina} \& {Popov}(2024)}]{Afonina2024}
{Afonina}, M.~D. \& {Popov}, S.~B. 2024, Universe, 10, 205

\bibitem[{{Arnaud}(1996)}]{Arnaud1996}
{Arnaud}, K.~A. 1996, in Astronomical Society of the Pacific Conference Series,
  Vol. 101, Astronomical Data Analysis Software and Systems V, ed. G.~H.
  {Jacoby} \& J.~{Barnes}, 17

\bibitem[{{Avakyan} {et~al.}(2023){Avakyan}, {Neumann}, {Zainab}, {Doroshenko},
  {Wilms}, \& {Santangelo}}]{Avakyan2023}
{Avakyan}, A., {Neumann}, M., {Zainab}, A., {et~al.} 2023, \aap, 675, A199

\bibitem[{{Avni}(1976)}]{Avni1976}
{Avni}, Y. 1976, \apj, 210, 642

\bibitem[{{Barlow} {et~al.}(2006){Barlow}, {Knigge}, {Bird}, {J Dean}, {Clark},
  {Hill}, {Molina}, \& {Sguera}}]{Barlow2006}
{Barlow}, E.~J., {Knigge}, C., {Bird}, A.~J., {et~al.} 2006, \mnras, 372, 224

\bibitem[{{Bird} {et~al.}(2006){Bird}, {Barlow}, {Bassani}, {Bazzano},
  {B{\'e}langer}, {Bodaghee}, {Capitanio}, {Dean}, {Fiocchi}, {Hill}, {Lebrun},
  {Malizia}, {Mas-Hesse}, {Molina}, {Moran}, {Renaud}, {Sguera}, {Shaw},
  {Stephen}, {Terrier}, {Ubertini}, {Walter}, {Willis}, \&
  {Winkler}}]{Bird2006}
{Bird}, A.~J., {Barlow}, E.~J., {Bassani}, L., {et~al.} 2006, \apj, 636, 765

\bibitem[{{Bird} {et~al.}(2010){Bird}, {Bazzano}, {Bassani}, {Capitanio},
  {Fiocchi}, {Hill}, {Malizia}, {McBride}, {Scaringi}, {Sguera}, {Stephen},
  {Ubertini}, {Dean}, {Lebrun}, {Terrier}, {Renaud}, {Mattana}, {G{\"o}tz},
  {Rodriguez}, {Belanger}, {Walter}, \& {Winkler}}]{Bird2010}
{Bird}, A.~J., {Bazzano}, A., {Bassani}, L., {et~al.} 2010, \apjs, 186, 1

\bibitem[{{Bird} {et~al.}(2016){Bird}, {Bazzano}, {Malizia}, {Fiocchi},
  {Sguera}, {Bassani}, {Hill}, {Ubertini}, \& {Winkler}}]{Bird2016}
{Bird}, A.~J., {Bazzano}, A., {Malizia}, A., {et~al.} 2016, \apjs, 223, 15

\bibitem[{{Bird} {et~al.}(2007){Bird}, {Malizia}, {Bazzano}, {Barlow},
  {Bassani}, {Hill}, {B{\'e}langer}, {Capitanio}, {Clark}, {Dean}, {Fiocchi},
  {G{\"o}tz}, {Lebrun}, {Molina}, {Produit}, {Renaud}, {Sguera}, {Stephen},
  {Terrier}, {Ubertini}, {Walter}, {Winkler}, \& {Zurita}}]{Bird2007}
{Bird}, A.~J., {Malizia}, A., {Bazzano}, A., {et~al.} 2007, \apjs, 170, 175

\bibitem[{{Brunschweiger} {et~al.}(2009){Brunschweiger}, {Greiner}, {Ajello},
  \& {Osborne}}]{Brunschweiger2009}
{Brunschweiger}, J., {Greiner}, J., {Ajello}, M., \& {Osborne}, J. 2009, \aap,
  496, 121

\bibitem[{{Campana} {et~al.}(2002){Campana}, {Stella}, {Israel}, {Moretti},
  {Parmar}, \& {Orlandini}}]{Campana2002}
{Campana}, S., {Stella}, L., {Israel}, G.~L., {et~al.} 2002, \apj, 580, 389

\bibitem[{{Chakrabarty} {et~al.}(2018{\natexlab{a}}){Chakrabarty}, {Jonker}, \&
  {Markwardt}}]{Chakrabarty2018a}
{Chakrabarty}, D., {Jonker}, P.~G., \& {Markwardt}, C.~B. 2018{\natexlab{a}},
  The Astronomer's Telegram, 11990, 1

\bibitem[{{Chakrabarty} {et~al.}(2018{\natexlab{b}}){Chakrabarty}, {Jonker}, \&
  {Markwardt}}]{Chakrabarty2018b}
{Chakrabarty}, D., {Jonker}, P.~G., \& {Markwardt}, C.~B. 2018{\natexlab{b}},
  The Astronomer's Telegram, 11991, 1

\bibitem[{{Chenevez} {et~al.}(2018){Chenevez}, {Jaisawal}, {Kuulkers},
  {Sanchez-Fernandez}, {Bazzano}, {Paizis}, {Bodaghee}, {Brandt}, {Lund},
  {Bozzo}, \& {Ferrigno}}]{Chenevez2018}
{Chenevez}, J., {Jaisawal}, G., {Kuulkers}, E., {et~al.} 2018, The Astronomer's
  Telegram, 11952, 1

\bibitem[{{de Martino} {et~al.}(2020){de Martino}, {Bernardini}, {Mukai},
  {Falanga}, \& {Masetti}}]{demartino2020}
{de Martino}, D., {Bernardini}, F., {Mukai}, K., {Falanga}, M., \& {Masetti},
  N. 2020, Advances in Space Research, 66, 1209

\bibitem[{{Doroshenko}(2024)}]{Doroshenko2024}
{Doroshenko}, V. 2024, arXiv e-prints, arXiv:2403.03127

\bibitem[{{Doroshenko} {et~al.}(2020){Doroshenko}, {Suleimanov}, {Tsygankov},
  {M{\"o}nkk{\"o}nen}, {Ji}, \& {Santangelo}}]{Doroshenko2020}
{Doroshenko}, V., {Suleimanov}, V., {Tsygankov}, S., {et~al.} 2020, \aap, 643,
  A62

\bibitem[{{Ferrigno} {et~al.}(2022){Ferrigno}, {Bozzo}, \&
  {Romano}}]{Ferrigno2022}
{Ferrigno}, C., {Bozzo}, E., \& {Romano}, P. 2022, \aap, 664, A99

\bibitem[{{Ferrigno} {et~al.}(2019){Ferrigno}, {Bozzo}, {Sanna}, {Jaisawal},
  {Girard}, {Di Salvo}, \& {Burderi}}]{Ferrigno2019}
{Ferrigno}, C., {Bozzo}, E., {Sanna}, A., {et~al.} 2019, \aap, 624, A142

\bibitem[{{Gosling} {et~al.}(2007){Gosling}, {Bandyopadhyay}, {Miller-Jones},
  \& {Farrell}}]{Gosling2007}
{Gosling}, A.~J., {Bandyopadhyay}, R.~M., {Miller-Jones}, J.~C.~A., \&
  {Farrell}, S.~A. 2007, \mnras, 380, 1511

\bibitem[{{Jansen} {et~al.}(2001){Jansen}, {Lumb}, {Altieri}, {Clavel}, {Ehle},
  {Erd}, {Gabriel}, {Guainazzi}, {Gondoin}, {Much}, {Munoz}, {Santos},
  {Schartel}, {Texier}, \& {Vacanti}}]{Jansen2001}
{Jansen}, F., {Lumb}, D., {Altieri}, B., {et~al.} 2001, \aap, 365, L1

\bibitem[{{Kretschmar} {et~al.}(2019){Kretschmar}, {F{\"u}rst}, {Sidoli},
  {Bozzo}, {Alfonso-Garz{\'o}n}, {Bodaghee}, {Chaty}, {Chernyakova},
  {Ferrigno}, {Manousakis}, {Negueruela}, {Postnov}, {Paizis}, {Reig},
  {Rodes-Roca}, {Tsygankov}, {Bird}, {Bissinger n{\'e} K{\"u}hnel}, {Blay},
  {Caballero}, {Coe}, {Domingo}, {Doroshenko}, {Ducci}, {Falanga}, {Grebenev},
  {Grinberg}, {Hemphill}, {Kreykenbohm}, {Kreykenbohm n{\'e} Fritz}, {Li},
  {Lutovinov}, {Mart{\'\i}nez-N{\'u}{\~n}ez}, {Mas-Hesse}, {Masetti},
  {McBride}, {Neronov}, {Pottschmidt}, {Rodriguez}, {Romano}, {Rothschild},
  {Santangelo}, {Sguera}, {Staubert}, {Tomsick}, {Torrej{\'o}n}, {Torres},
  {Walter}, {Wilms}, {Wilson-Hodge}, \& {Zhang}}]{Kretschmar2019}
{Kretschmar}, P., {F{\"u}rst}, F., {Sidoli}, L., {et~al.} 2019, \nar, 86,
  101546

\bibitem[{{Krivonos} {et~al.}(2007){Krivonos}, {Revnivtsev}, {Lutovinov},
  {Sazonov}, {Churazov}, \& {Sunyaev}}]{Krivonos2007}
{Krivonos}, R., {Revnivtsev}, M., {Lutovinov}, A., {et~al.} 2007, \aap, 475,
  775

\bibitem[{{Lucas} {et~al.}(2008){Lucas}, {Hoare}, {Longmore}, {Schr{\"o}der},
  {Davis}, {Adamson}, {Bandyopadhyay}, {de Grijs}, {Smith}, {Gosling},
  {Mitchison}, {G{\'a}sp{\'a}r}, {Coe}, {Tamura}, {Parker}, {Irwin}, {Hambly},
  {Bryant}, {Collins}, {Cross}, {Evans}, {Gonzalez-Solares}, {Hodgkin},
  {Lewis}, {Read}, {Riello}, {Sutorius}, {Lawrence}, {Drew}, {Dye}, \&
  {Thompson}}]{Lucas2008}
{Lucas}, P.~W., {Hoare}, M.~G., {Longmore}, A., {et~al.} 2008, \mnras, 391, 136

\bibitem[{{Mart{\'\i}nez-N{\'u}{\~n}ez}
  {et~al.}(2017){Mart{\'\i}nez-N{\'u}{\~n}ez}, {Kretschmar}, {Bozzo},
  {Oskinova}, {Puls}, {Sidoli}, {Sundqvist}, {Blay}, {Falanga}, {F{\"u}rst},
  {G{\'\i}menez-Garc{\'\i}a}, {Kreykenbohm}, {K{\"u}hnel}, {Sander},
  {Torrej{\'o}n}, \& {Wilms}}]{Martinez2017}
{Mart{\'\i}nez-N{\'u}{\~n}ez}, S., {Kretschmar}, P., {Bozzo}, E., {et~al.}
  2017, \ssr, 212, 59

\bibitem[{{Masetti} {et~al.}(2018){Masetti}, {Ferreira}, {Saito}, {Kammers}, \&
  {Minniti}}]{Masetti2018}
{Masetti}, N., {Ferreira}, T.~S., {Saito}, R.~K., {Kammers}, R., \& {Minniti},
  D. 2018, The Astronomer's Telegram, 11992, 1

\bibitem[{{Masetti} {et~al.}(2006){Masetti}, {Orlandini}, {Palazzi}, {Amati},
  \& {Frontera}}]{Masetti2006}
{Masetti}, N., {Orlandini}, M., {Palazzi}, E., {Amati}, L., \& {Frontera}, F.
  2006, \aap, 453, 295

\bibitem[{McCollum {et~al.}(2018)McCollum, Laine, \& McCollum}]{McCollum2018}
McCollum, B., Laine, S., \& McCollum, M. 2018, Research Notes of the AAS, 2,
  193

\bibitem[{{Negueruela} {et~al.}(2006){Negueruela}, {Smith}, {Reig}, {Chaty}, \&
  {Torrej{\'o}n}}]{Negueruela2006}
{Negueruela}, I., {Smith}, D.~M., {Reig}, P., {Chaty}, S., \& {Torrej{\'o}n},
  J.~M. 2006, in Proc. of the ``The X-ray Universe 2005'', 26-30 September
  2005, El Escorial, Madrid, Spain. Ed. by A. Wilson. ESA SP-604, Volume 1,
  Noordwijk: ESA Pub. Division, ISBN 92-9092-915-4, 2006

\bibitem[{{Ochsenbein} {et~al.}(2000){Ochsenbein}, {Bauer}, \&
  {Marcout}}]{vizier2000}
{Ochsenbein}, F., {Bauer}, P., \& {Marcout}, J. 2000, \aaps, 143, 23

\bibitem[{{Ransom} {et~al.}(2002){Ransom}, {Eikenberry}, \&
  {Middleditch}}]{Ransom2002}
{Ransom}, S.~M., {Eikenberry}, S.~S., \& {Middleditch}, J. 2002, \aj, 124, 1788

\bibitem[{{Revnivtsev} {et~al.}(2008){Revnivtsev}, {Sazonov}, {Krivonos},
  {Ritter}, \& {Sunyaev}}]{Revnivtsev2008}
{Revnivtsev}, M., {Sazonov}, S., {Krivonos}, R., {Ritter}, H., \& {Sunyaev}, R.
  2008, \aap, 489, 1121

\bibitem[{{Sguera} {et~al.}(2005){Sguera}, {Barlow}, {Bird}, {Clark}, {Dean},
  {Hill}, {Moran}, {Shaw}, {Willis}, {Bazzano}, {Ubertini}, \&
  {Malizia}}]{Sguera2005}
{Sguera}, V., {Barlow}, E.~J., {Bird}, A.~J., {et~al.} 2005, \aap, 444, 221

\bibitem[{{Sguera} {et~al.}(2006){Sguera}, {Bazzano}, {Bird}, {Dean},
  {Ubertini}, {Barlow}, {Bassani}, {Clark}, {Hill}, {Malizia}, {Molina}, \&
  {Stephen}}]{Sguera2006}
{Sguera}, V., {Bazzano}, A., {Bird}, A.~J., {et~al.} 2006, \apj, 646, 452

\bibitem[{{Shakura} {et~al.}(2012){Shakura}, {Postnov}, {Kochetkova}, \&
  {Hjalmarsdotter}}]{Shakura2012}
{Shakura}, N., {Postnov}, K., {Kochetkova}, A., \& {Hjalmarsdotter}, L. 2012,
  \mnras, 420, 216

\bibitem[{{Shakura} {et~al.}(2014){Shakura}, {Postnov}, {Sidoli}, \&
  {Paizis}}]{Shakura2014}
{Shakura}, N., {Postnov}, K., {Sidoli}, L., \& {Paizis}, A. 2014, \mnras, 442,
  2325

\bibitem[{{Sidoli} \& {Paizis}(2018)}]{Sidoli2018}
{Sidoli}, L. \& {Paizis}, A. 2018, \mnras, 481, 2779

\bibitem[{{Sidoli} {et~al.}(2023){Sidoli}, {Ponti}, {Sguera}, \&
  {Esposito}}]{Sidoli2023}
{Sidoli}, L., {Ponti}, G., {Sguera}, V., \& {Esposito}, P. 2023, \aap, 671,
  A150

\bibitem[{{Str{\"u}der} {et~al.}(2001){Str{\"u}der}, {Briel}, {Dennerl},
  {Hartmann}, {Kendziorra}, {Meidinger}, {Pfeffermann}, {Reppin}, \& {et
  al.}}]{Struder2001}
{Str{\"u}der}, L., {Briel}, U., {Dennerl}, K., {et~al.} 2001, \aap, 365, L18

\bibitem[{{Suleimanov} {et~al.}(2022){Suleimanov}, {Doroshenko}, \&
  {Werner}}]{Suleimanov2022}
{Suleimanov}, V.~F., {Doroshenko}, V., \& {Werner}, K. 2022, \mnras, 511, 4937

\bibitem[{{Tomsick} {et~al.}(2009){Tomsick}, {Chaty}, {Rodriguez}, {Walter}, \&
  {Kaaret}}]{Tomsick2009}
{Tomsick}, J.~A., {Chaty}, S., {Rodriguez}, J., {Walter}, R., \& {Kaaret}, P.
  2009, \apj, 701, 811

\bibitem[{{Turner} {et~al.}(2001){Turner}, {Abbey}, {Arnaud}, {Balasini},
  {Barbera}, {Belsole}, {Bennie}, {Bernard}, \& {et al.}}]{Turner2001}
{Turner}, M.~J.~L., {Abbey}, A., {Arnaud}, M., {et~al.} 2001, \aap, 365, L27

\bibitem[{{Verner} {et~al.}(1996){Verner}, {Ferland}, {Korista}, \&
  {Yakovlev}}]{Verner1996}
{Verner}, D.~A., {Ferland}, G.~J., {Korista}, K.~T., \& {Yakovlev}, D.~G. 1996,
  \apj, 465, 487

\bibitem[{{Wilms} {et~al.}(2000){Wilms}, {Allen}, \& {McCray}}]{Wilms2000}
{Wilms}, J., {Allen}, A., \& {McCray}, R. 2000, \apj, 542, 914

\bibitem[{{Xu} {et~al.}(2016){Xu}, {Wang}, \& {Li}}]{xu2016}
{Xu}, X.-j., {Wang}, Q.~D., \& {Li}, X.-D. 2016, \apj, 818, 136

\bibitem[{{Yungelson} {et~al.}(2019){Yungelson}, {Kuranov}, \&
  {Postnov}}]{Yungelson2019}
{Yungelson}, L.~R., {Kuranov}, A.~G., \& {Postnov}, K.~A. 2019, \mnras, 485,
  851

\bibitem[{{Zolotukhin} \& {Revnivtsev}(2011)}]{Zolotukhin2011}
{Zolotukhin}, I.~Y. \& {Revnivtsev}, M.~G. 2011, \mnras, 411, 620

\end{thebibliography}

\end{document}